\documentclass[pra, showpacs, twocolumn, floatfix, superscriptaddress]{revtex4}

\usepackage{graphicx}
\usepackage{amsmath, amsfonts, amssymb, bm}
\usepackage{nicefrac}
\usepackage{longtable}

\usepackage[]{color}
\usepackage{amstext}
\usepackage{mathrsfs}
\usepackage[]{psfrag}

\usepackage[caption=false]{subfig}

\usepackage[normalem]{ulem}

\renewcommand{\vec}{\bm}
\newcommand{\dirac}{\displaystyle{\not}}
\newcommand{\e}{\text{e}}

\newcommand{\ii}{\text{i}}
\newcommand{\dif}{\text{d}}
\psfragscanoff
\providecommand{\abs}[1]{\lvert#1\rvert}

\begin{document}
\title{Particle production reactions in laser-boosted lepton collisions}
\author{Sarah J. M\"uller}
\affiliation{Max-Planck-Institut f\"ur Kernphysik, Saupfercheckweg 1, D-69117 Heidelberg, Germany}
\author{Christoph H. Keitel}
\affiliation{Max-Planck-Institut f\"ur Kernphysik, Saupfercheckweg 1, D-69117 Heidelberg, Germany}
\author{Carsten M\"uller}
\affiliation{Max-Planck-Institut f\"ur Kernphysik, Saupfercheckweg 1, D-69117 Heidelberg, Germany}\affiliation{Institut f\"ur Theoretische Physik I, Heinrich-Heine-Universit\"at D\"usseldorf, Universit\"atsstra\ss e 1, 40225 D\"usseldorf, Germany}\email{Carsten.Mueller@tp1.uni-duesseldorf.de}

\begin{abstract}%
The need for ever higher energies in lepton colliders gives rise to the investigation of new accelerator schemes for elementary particle physics experiments. One perceivable way to increase the collision energy would be to combine conventional lepton acceleration with strong laser fields, making use of the momentum boost a charged particle experiences inside a plane electromagnetic wave. As an example for a process taking place in such a laser-boosted collision, Higgs boson creation is studied in detail. We further discuss other possible particle production processes that could be implemented in such a collider scheme and specify the required technical demands.
\end{abstract}

\maketitle

\section{Introduction}

\begin{figure}
  \includegraphics[width=0.25\textwidth]{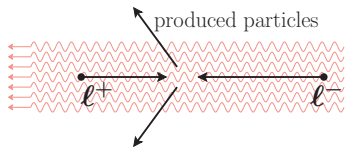}
\caption{(Color online) Schematic view of the considered collision setup. Preaccelerated leptons and antileptons are superimposed by a strong laser field (red wiggled lines), which further boosts the collision energy.}\label{schemabild}
\end{figure}

Experiments of high-energy physics require increasingly powerful and highly energetic particle colliders. Conventional accelerators, however, face certain limitations with regards to damping, size and cost. For example, the Large Electron-Positron Collider (LEP), the most powerful lepton accelerator in history, could not reach sufficiently high energies for Higgs boson creation to become possible in electron-positron collisons. Damping effects due to its circular shape have rendered further increase of the collision energy impossible and so the LEP was replaced by the Large Hadron Collider (LHC) which, recently, succeeded in the discovery of the Higgs boson \cite{Higgs,Higgs2}. Still, Higgs boson production in lepton collisions is an important process to investigate, since certain properties and couplings of the Higgs boson cannot be examined in hadron collisions \cite{Higgs3}. But the required collision energies can, by conventional means, only be obtained in even larger circular or newly built linear colliders, both of which are very cost intensive. For these reasons, alternative acceleration methods are under active scrutiny. \\

In view of the fast development of power and intensity of modern laser facilities, laser acceleration becomes increasingly interesting for high-energy physics processes. The most intense laser facility to date is the HERCULES laser in Michigan, reaching intensities over $10^{22}$ W/cm$^2$ \cite{Yanovsky}. The record in laser pulse energy is currently held by the National Ignition Facility (NIF) in Livermore, California, with peak powers up to 500 TW and pulse energies up to 2 MJ \cite{NIF}. The most powerful lasers, such as the Vulcan Laser in the United Kingdom and the Texas Petawatt Laser, reach pulse powers in the petawatt regime \cite{petawattlaser1, petawattlaser2}. In Europe, the future Extreme Light Infrastructure (ELI) currently under construction is bound to reach intensities over $10^{25}$ W/cm$^2$ and pulse powers in the multi-10 PW regime \cite{ELI}. In Russia, the Exawatt Center for Extreme Light Studies (XCELS) project has similar goals \cite{XCELS}. Laser facilities with such high intensities and pulse powers offer broad opportunities with regard to strong-field quantum electrodynamics (QED) \cite{kaminskireview, dipiazzareview}, laser-based high-energy physics \cite{hep1,hep2,hep3,hep4}, and studies of physics beyond the standard model \cite{beyond1,beyond2,beyond3,beyond4, beyond5, beyond6, beyond7}.\\

Concerning alternative particle acceleration schemes, one very promising ansatz involves laser acceleration based on laser-plasma interaction. The so-called laser wakefield acceleration (LFWA) has successfully been realized in experiments, accelerating electrons from solid targets to a few GeV \cite{Leemans,Wang}, producing highly relativistic positron beams with energies up to the 100 MeV regime \cite{Chen,Chen2, Sarri}, or increasing the energy of parts of a preaccelerated electron beam \cite{Blumenfeld} along very small accelerator extensions. In the latter case, the plasma wave was induced by the initial particle beam itself instead of an external laser wave. An LWFA-based TeV-lepton collider has been proposed \cite{Leemans2}.\\

Another acceleration scheme involving strong laser fields is laser-particle acceleration directly in the vacuum \cite{Chang, Galow}. Here, one makes use of the fact that charged particles, while traveling in a strong electromagnetic wave, may temporarily gain vast amounts of energy. The applicability of this mechanism to high-energy physics experiments is studied in \cite{McDonald, KarenEPL, Eminov}. Relativistic electron-positron pairs have been produced in high-energy electron-laser collisions \cite{Burke} and laser-accelerated electron beams were utilized to study Thomson scattering \cite{Cowan}. \\

In this paper, we study various processes of particle physics in a collider scheme where lepton and antilepton beams are preaccelerated by conventional methods and superimposed by a strong laser field prior to the collision so that the latter takes place inside the laser wave and the particles are further accelerated to higher collision energies (Fig. \ref{schemabild}). Based on this concept, which has first been introduced in \cite{McDonald}, we have recently presented in \cite{plb} an investigation of Higgs boson production in lepton collisions which were boosted by a linearly polarized laser field. We now expand our investigation of this exemplary electroweak process by analyzing its dependence on the applied laser field amplitude, frequency and polarization and the incident lepton energy. Additionally, we study further particle creation processes: as an example of a purely QED process, we examine the production of muon-antimuon pairs in laser-boosted electron-positron collisions and finally, we discuss resonant hadron production. \\

The paper is organized as follows. We will first specify the units system and the notation used and briefly introduce the concept of ``laser-dressing'' of elementary particles in plane electromagnetic waves. In Sec.~\ref{HiggsSection}, we present the calculation of the process $\ell^+\ell^-\to HZ$ inside a laser wave, first in a circularly polarized field, and then in a linearly polarized one. We then specify the properties of the laser pulse required for the parameter sets to which we give numerical results. In Sec.~\ref{otherprocesses}, we first present in detail the corresponding calculation and numerical results for the process $e^+e^-\to \mu^+\mu^-$ for both considered laser polarizations and, again, specify the experimental demands. After that, we discuss the laser parameters that would be necessary for the resonant production of neutral pions, $\Phi$ mesons and $J/\Psi$ mesons. In Sec.~\ref{summarySection}, we summarize our results.

\subsection{Notation}

Throughout our calculations, we use a natural units system with $\hbar=c=4\pi\varepsilon_0=1$ and $e=\sqrt{\alpha}$ with the fine structure constant $\alpha$ \cite{PeskinSchroeder}, and the metric tensor in its form $g^{\mu\nu}=\text{diag}(1,-1,-1,-1)$. The $\gamma$ matrices are employed in the Dirac representation,
\begin{equation}
 \gamma^0=\begin{pmatrix}
                 \bm{1}&0\\
		0&-\bm{1}
                \end{pmatrix}\,,\qquad \gamma^i=\begin{pmatrix}
                 0&\sigma^i\\
		-\sigma^i&0
                \end{pmatrix}\,
\end{equation}
with the Pauli matrices $\sigma_i$. The fifth $\gamma$ matrix is given by $\gamma^5=i\gamma^0\gamma^1\gamma^2\gamma^3$ \cite{BjorkenDrell}. We employ Feynman slash notation for four-products of four-vectors with $\gamma$ matrices, $\gamma^\mu a_\mu=\dirac a$. Four-products of four-vectors $a$, $b$ are written as $(ab)$. Laser potentials $A^\mu$ are used in the Lorenz gauge, i.e. $(\partial A)=\partial_\mu A^\mu = 0$ and $A^0=0$ \cite{BerestetskiiQED}.

\subsection{Leptons in Plane Electromagnetic Waves}\label{Volkov_section}
We will now briefly summarize the relevant equations for lepton motion in plane electromagnetic waves. A detailed derivation can be found in \cite{Itzykson, BerestetskiiQED}. The state $\psi$ of a lepton with charge $-e$ and mass $m$ traveling in an electromagnetic field with the four-potential $A$ must solve the Dirac equation

\begin{equation}\label{DiracEquationInLaserField}
( \ii\dirac\partial+e\dirac A -m)\psi=0\,.
\end{equation}
The solutions $\psi$ for a plane electromagnetic wave with wave vector $k$ and frequency $\omega$ have first been calculated by Volkov in 1935 \cite{VolkovRef} and are therefore called the Volkov states. Their general form is given by
\begin{equation}\label{VolkovStateAllgLepton}
 \psi_-  = N_p\left(1-\frac{e\dirac k\dirac A}{2(kp)}\right)u_-(p,s)\e^{\ii S^{-}}
\end{equation}
for leptons and 
\begin{equation}\label{VolkovStateAllgAntilepton}
  \psi_+ = N_p\left(1+\frac{e\dirac k\dirac A}{2(kp)}\right)u_+(p,s)\e^{\ii S^{+}}
\end{equation}
for antileptons. The spinors $u_\pm$ are free Dirac spinors, fulfilling the normalization $\overline{u}u=2m$, $N_p$ is a normalization, and $S^\pm$ is given by
\begin{equation}
 S^\pm = \pm (px) +\frac{e}{(kp)}\int^\kappa \dif \tilde\kappa \left[\left(pA(\tilde\kappa)\right) \mp \frac{e}{2}A^2(\tilde\kappa)\right]\,.
\end{equation}
It corresponds to the classical action of the (anti)lepton in the laser field.\\
The time average of the current density $j^\mu$ corresponding to the Volkov state \eqref{VolkovStateAllgLepton} is given by
\begin{equation}
 \overline{j^\mu}=2\abs{N_p}^2\left(p^\mu - \frac{e^2\overline{A^2}}{2(kp)}k^\mu\right)\,,
\end{equation}
where we used $\overline{u}\gamma^\mu u = 2 p^\mu$. We can thus associate the lepton motion with an {effective} or {\em laser-dressed} momentum
\begin{equation}\label{qm_qdef}
 q^\mu:=p^\mu - \frac{e^2\overline{A^2}}{2(kp)}k^\mu = p^\mu+\frac{m^2\xi^2}{2(kp)}k^\mu
\end{equation}
with which we can choose the normalization constant to be $N_p=\sqrt{1/2q^0}$ so that the 0-component of the current density is normalized to $\overline{j^0}=1$. In Eq. \eqref{qm_qdef}, we introduced the laser intensity parameter 
\begin{equation}\label{xidefallg}
 \xi := \frac{\abs{e}}{m}\sqrt{-\overline{A^2}}\,
\end{equation}
which is a dimensionless measure for the impact of the laser wave on the lepton. Corresponding to the effective momenta $q$, we can introduce an effective mass $m_*=\sqrt{1+\xi^2}m$, for which $(q)^2=m_*^2$.\\
The laser-dressed momenta $q$ from Eq. \eqref{qm_qdef} can be much larger than the free lepton momenta $p$: if the leptons are traveling with a relativistic velocity in the direction of the laser propagation, the four-product $(kp)$ can become very small and thus the field-dependent summand containing the laser impact may become very large. {In the case where a lepton and an antilepton beam are traveling perfectly collinear with the laser propagation direction, the laser-dressed energies amount to}
\begin{equation}\label{q0eq}
 q^0\approx\begin{cases} (1+\xi^2)p^0 & \text{coprop. $\ell^-$} \\ p^0 & \text{counterprop. $\ell^+$}\end{cases}\,.
\end{equation}
{While, depending on the size of the laser intensity parameter $\xi$, this may result in a strong asymmetry in the collision geometry, it may also lead to a substantial increase of the collision energy.}

\section{Higgs Boson Creation}\label{HiggsSection}
\begin{figure}
 \includegraphics[width=0.25\textwidth]{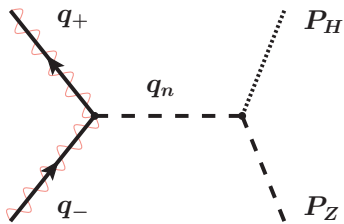}
\caption{(Color online) Feynmann diagram of the most important Higgs boson creation process in a lepton collision, $\ell^+\ell^-\to HZ$. The wiggled lines indicate that the (electrically charged) leptons with quasimomenta $q_\pm$ interact with a superimposed laser field.}\label{HiggsFeyn}
\end{figure}

\begin{figure}
 \includegraphics[width=.45\textwidth]{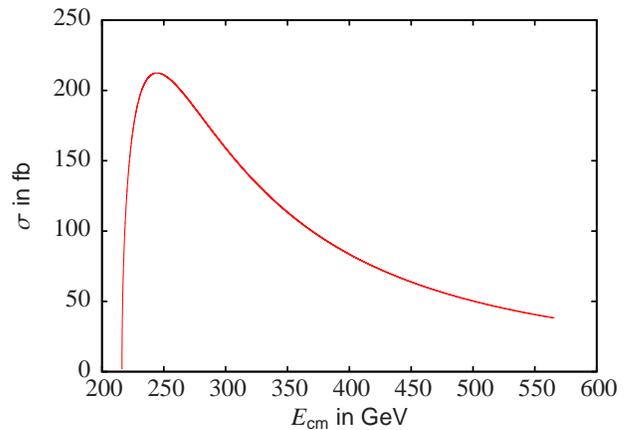}
\caption{(Color online) Dependence of the field free cross section \eqref{sigma_ff} on the collision energy $\sqrt{s}$ for a Higgs boson with mass $M_H=125$ GeV in the process $\ell^+\ell^-\to HZ$. The cross section has a maximum of $\sigma_{\text{ff}}\approx212$ fb at $\sqrt{s}\approx244.5$ GeV.}\label{ff_sigplot}
\end{figure}

In lepton collisions, the most important Higgs boson creation mechanism is the associated production with a $Z$ boson via the process $\ell^+\ell^-\to Z^*\to ZH$ (cf. Fig. \ref{HiggsFeyn}). Without a laser field, the well-known transition amplitude for this process reads \cite{Donoghue}
\begin{align}\label{ampli_ff}
\mathcal S_{\text{ff}}=&\overline u_+\frac{-\ii g}{2\cos\theta_W}(g_V\gamma^\mu-g_A\gamma^\mu\gamma^5)u_-\nonumber\\
&\cdot\frac{-4\pi\ii g_{\mu\nu}+\frac{4\pi\ii q_\mu q_\nu}{M_Z^2}}{q^2-M_Z^2+\ii\Gamma M_Z}\cdot\frac{\ii gM_Zg^{\nu\rho}}{\cos\theta_W}\epsilon^*_\rho(P_Z)\,,
\end{align}
with the Weinberg angle $\theta_W$, the weak coupling constant $g=e/\sin\theta_W$, and the leptonic weak neutral coupling constants $g_V$ and $g_A$. The first line of Eq. \eqref{ampli_ff} describes the lepton-antilepton vertex and the second line contains the propagator of the virtual $Z$ boson as well as the outgoing particles' vertex. In the virtual $Z$ boson's propagator, the decay width $\Gamma$ of the $Z$ boson can be neglected due to the large collision energy involved. Further neglecting terms of the order $m/M_Z$, $m/M_H$, and $m/\sqrt{s}$, which is justified because the lepton mass $m$ is small compared to the large masses of the produced bosons $M_Z$ and $M_H$ as well as the large collision energy $\sqrt{s}$, this leads to the total cross section
\begin{align}\label{sigma_ff}
 \sigma_{\text{ff}} (\sqrt{s})=& \frac{\sqrt{\lambda(s,M_Z^2,M_H^2)}}{s}\frac{\pi\alpha^2(g_V^2+g_A^2)}{48s\cos^4\theta_W\sin^4\theta_W}\nonumber\\
 &\cdot\frac{12sM_Z^2+\lambda(s,M_Z^2,M_H^2)}{(s-M_Z^2)^2}\,
\end{align}
with 
\begin{equation}\label{lambdadef}
  \lambda(s,M_Z^2,M_H^2) := s^2 + M_Z^4 + M_H^4 -2sM_Z^2 -2sM_H^2 -2M_Z^2M_H^2\,.
\end{equation}
The total cross section (without a laser field) thus only depends on the collision energy $\sqrt{s}$ as shown in Fig. \ref{ff_sigplot}. Please note that it does not depend on the species of colliding leptons as long as their mass is well below the collision energy and masses of the produced bosons, i.e. Eq. \eqref{sigma_ff} holds for muon-antimuon collisions as well as electron-positron collisions.\\

Inside the laser field, the leptons are no longer described by free Dirac spinors $u_\pm$ as in Eq. \eqref{ampli_ff}, but the Volkov states from Sec.~\ref{Volkov_section} have to be used. With this, the transition amplitude reads, in position space,
\begin{align}\label{TransitionAmplitude}
\mathcal S=& \frac{-\ii g}{2\cos\theta_W}\iiint \overline \psi_+(x)\gamma_\mu (g_V-g_A\gamma_5)\psi_{-}(x) \nonumber\\
&\cdot  \frac{4\pi\e^{\ii q(x-y)}}{(2\pi)^4}\frac{-\ii g^{\mu\nu}+\frac{\ii q^\mu q^\nu}{M_Z^2}}{q^2-M_Z^2}\nonumber\\
&\cdot \frac{\ii gM_Zg_{\nu\rho}}{\cos\theta_W}\frac{1}{2\sqrt{E_ZE_H}}\epsilon^*_\rho(P_Z)\e^{\ii(P_Z+P_H)y}\dif^4q\dif^4x \dif^4y\,.
\end{align}

Since the Volkov states $\psi_\pm(x)$ depend on the laser potential $A$, the polarization of the laser field plays a role in the evaluation of the transition amplitude \eqref{TransitionAmplitude}. In the following, we show the analytical calculation as well as some numerical results for circular and linear polarization of the laser field.

\subsection{Circular Polarization}\label{HiggsCPsection}

We now consider the process $\ell^+\ell^-\to HZ$ inside a laser field with circular polarization, i.e. with the laser potential 
\begin{equation}
 A_c(\kappa) \, = \, a_1  \cos(\kappa)  +  a_2  \sin(\kappa)\,
\end{equation}
with $\kappa=(kx)$, $a_1=(0,a,0,0)$ and $a_2=(0,0,a,0)$, $a$ being the amplitude of the laser field. With this, the Volkov states from Eqs. \eqref{VolkovStateAllgLepton} and \eqref{VolkovStateAllgAntilepton} become
\begin{align}\label{VolkovstatesCircularpol}
 \psi_{\pm}(x)= &\sqrt{\frac{1}{2q^0_\pm}}\left(1\pm\frac{e\dirac k(\dirac a_1\cos\kappa+\dirac a_2\sin\kappa)}{2(kp_\pm)}\right)u_\pm\nonumber\\
&\times\e^{\pm\ii(q_\pm x) } \e^{ \frac{\ii e}{(kp_\pm)}\bigl((p_\pm a_1)\sin\kappa - (p_\pm a_2)\cos\kappa\bigr)}\,
 \end{align}
 with the laser-dressed momenta 
 \begin{equation}
   q_\pm^\mu = p_\pm^\mu + \xi_c^2 \frac{m^2}{2(kp_\pm)}k^\mu\,
 \end{equation}
containing the laser intensity parameter
\begin{equation}
 \xi_c=\frac{ea}{m}\,.
\end{equation}

With the abbreviation $\Gamma_\mu:=\gamma_\mu(g_V-g_A\gamma_5)$, we can now investigate the leptonic current {$\mathcal{J}_\mu=\int\dif^4x\,\overline\psi_+\Gamma_\mu\psi_-$}. Written out, it reads
\begin{align}
 \mathcal{J}_\mu&=\frac{1}{2\sqrt{q_+^0q_-^0}}\nonumber\\
 &\quad\times\int\dif^4x\,\overline{u}_{+}\left[\Bigl(\gamma^\mu-\frac{e^2a^2k_\mu}{2(kp_+)(kp_-)}\dirac k\Bigr)(g_V-g_A\gamma_5)\right. \nonumber\\
&\qquad+\frac{e}{2} \Bigl(\bigl(\frac{1}{(kp_+)}\dirac a_1\dirac k\Gamma_\mu - \frac{1}{(kp_-)}\Gamma_\mu\dirac k\dirac a_1\bigr)\cos\kappa\nonumber\Bigr.\\
&\qquad\,\left.\Bigl. + \bigl(\frac{1}{(kp_+)}\dirac a_2\dirac k\Gamma_\mu-\frac{1}{(kp_-)}\Gamma_\mu\dirac k\dirac a_2\bigr)\sin\kappa\Bigr)\right]u_{-}\nonumber\\
&\quad\times \e^{-\ii\bigl(q_++q_-\bigr)x}\times\e^{-\ii\bigl(\alpha_1\sin\kappa - \alpha_2\cos\kappa\bigr)}\,
\label{ElectronCurrentExpansion_cp}
\end{align}

with
\begin{equation}
 \alpha_j:=\frac{e(p_+a_j)}{(kp_+)}-\frac{e(p_-a_j)}{(kp_-)}\,.\label{betaDef}
\end{equation}

The second exponential function in Eq. \eqref{ElectronCurrentExpansion_cp}, as well as its products with $\cos\kappa$ and $\sin\kappa$, can be expanded in a Fourier series, as is standard procedure in QED
 processes \cite{Ritus},
 
\begin{align}\label{FourierElectrons_cp}
 f(\kappa):=\e^{-\ii\bigl(\alpha_1\sin\kappa - \alpha_2\cos\kappa\bigr)}&=\sum_{n=-\infty}^\infty b_n \e^{-\ii n\kappa}\,,\nonumber\\
  \cos(\kappa) \,f(\kappa)&=\sum_{n=-\infty}^\infty c_n \e^{-\ii n\kappa}\,,\nonumber\\
  \sin(\kappa) \,f(\kappa)&=\sum_{n=-\infty}^\infty d_n \e^{-\ii n\kappa}\,
\end{align}
with the coefficients
\begin{align}\label{FourierCoeffElectrons_cp}
  b_n&=J_n(\overline{\alpha})\e^{\ii n\kappa_0}\,,\nonumber\\
  c_n&=\frac{1}{2}\biggl(J_{n+1}(\overline{\alpha})\,\e^{\ii(n+1)\kappa_0}+J_{n-1}(\overline{\alpha})\,\e^{\ii(n-1)\kappa_0} \biggr)\,,\nonumber\\
  d_n&=\frac{1}{2\ii}\biggl(J_{n+1}(\overline{\alpha})\,\e^{\ii(n+1)\kappa_0}-J_{n-1}(\overline{\alpha})\,\e^{\ii(n-1)\kappa_0} \biggr)\,
\end{align}
containing regular cylindrical Bessel functions $J_n$ of integer order $n$ \cite{Abramowitz}. Their argument is 
\begin{equation}\label{BessArg_cp}
 \overline\alpha=\sqrt{\alpha_1^2+\alpha_2^2}
\end{equation}
and the angle $\kappa_0$ is given by
\begin{equation}\label{kappa_0_cp}
 \cos\kappa_0=\frac{\alpha_1}{\overline\alpha}\,,\quad\sin\kappa_0=\frac{\alpha_2}{\overline\alpha}\,.
\end{equation}
Thus, we can write the leptonic current as 
\begin{equation}\label{lepcurrent}
 \mathcal{J}_\mu=\frac{1}{2\sqrt{q_+^0q_-^0}} \sum_{n=-\infty}^\infty\int\dif^4x\mathcal{M}_\mu^n\cdot\e^{-\ii\bigl(q_++q_-\bigr)x}\cdot\e^{-\ii n(kx)}\,
\end{equation}
containing the spinor-matrix products $\mathcal{M}_\mu^n$. This allows us to perform the integrals in Eq. \eqref{TransitionAmplitude}, leading to the transition amplitude
\begin{align}\label{S_n_cp}
 \mathcal S =\sum_n& \frac{(2\pi)^5}{4\sqrt{q_+^0q_-^0E_ZE_H}}\frac{g^2M_Z}{\cos^2\theta_W}\nonumber\\
 &\times\mathcal M_\mu^n\frac{-ig^{\mu\nu}+\frac{\ii q_n^\mu q_n^\nu}{M_Z^2}}{q_n^2-M_Z^2}g_{\nu\rho}\epsilon^*_\rho(P_Z)\delta(q_n-P_H-P_Z)\nonumber\\
=:\sum_n&\mathcal{S}_n\,,
\end{align}
which is a sum over {partial} transition amplitudes $\mathcal{S}_n$ depending on the order $n$ of the Bessel functions, which may be interpreted as the number of absorbed (if $n>0$) or emitted (if $n<0$) laser photons. The four-momentum of the virtual $Z$ boson, $q_n = q_++q_-+nk$, is now also dependent on the photon order $n$.\\

The cross section is derived from the transition amplitude via
\begin{equation} \label{totdiffsigma_cp_posspace_1}
 \dif^6\sigma = \frac{1}{4}\sum_{\text{pol.}}\sum_{\text{spins}}\frac{\abs{\mathcal S}^2}{\tau\abs{\vec j}}\frac{\dif^3\vec P_Z}{(2\pi)^3}\frac{\dif^3\vec P_H}{(2\pi)^3}\,
\end{equation}
with the unit time $\tau$ and the laser-dressed flux $\vec j$ of the incoming particles \cite{Berestetskii}
\begin{equation}\label{cp_leptonflux}
 \abs{\vec j}=\frac{\sqrt{(q_+q_-)^2-m_*^4}}{q_+^0q_-^0}\,.
\end{equation}
With the abbreviations $T_n:=\mathcal M_\mu^n(-g^{\mu\rho}+\frac{q_n^\mu q_n^\rho}{M_Z^2})\epsilon^*_\rho(P_Z)$ and $t_n:=\sum_{\text{pol.}}\sum_{\text{spins}} T_n T_n^\dagger$ containing lengthy traces over products of up to eight $\gamma$ matrices, squaring of the transition amplitude leads to 
\begin{align}\label{d6sigma_cp}
  \dif^6\sigma =& \sum_n \frac{ g^4M_Z^2}{64E_ZE_H\cos^4\theta_W\sqrt{(q_+q_-)^2-m_*^4}}\nonumber\\&\times\frac{t_n\delta(q_n-P_Z-P_H)}{(q_n^2-M_Z^2)^2}\dif^3\vec P_Z \dif^3\vec P_H\,.
 \end{align}
Integration over the $Z$ boson's momentum $\dif^3\vec P_Z$ eliminates three of the four dimensions of the $\delta$ function and leads to $\vec P_Z = \vec q_n-\vec P_H$. The remaining energy-conserving $\delta$ function $\delta(q_n^0-E_Z-E_H)$ is eliminated by first rewriting $\dif^3\vec P_H = \abs{\vec P_H}E_H\dif E_H\dif\Phi_H\dif\cos\Theta_H$ and then performing the integration over $\dif\cos \Theta_H$, leading to
\begin{align}\label{higgscos_cp}
 \cos\Theta_H=\frac{M_Z^2-M_H^2-q_n^2+2q_n^0E_H}{2\abs{\vec q_n}\abs{\vec P_H}}\,.
\end{align}
From the restriction that $\abs{\cos\Theta_H}\le1$ follow the integration limits 
\begin{align}\label{higgsenergylimits_cp}
 E_H^{\text{min}} = \frac{1}{2}(q_n^2+M_H^2-M_Z^2)\frac{q_n^0}{(q_n)^2} - \frac{\abs{\vec q_n}}{2(q_n)^2}\sqrt{\lambda(q_n^2, M_Z^2, M_H^2)}\nonumber\\
E_H^{\text{max}} = \frac{1}{2}(q_n^2+M_H^2-M_Z^2)\frac{q_n^0}{(q_n)^2} + \frac{\abs{\vec q_n}}{2(q_n)^2}\sqrt{\lambda(q_n^2, M_Z^2, M_H^2)}\,
\end{align}
for the integration over $\dif E_H$, and the integration over the Higgs boson's azimuth angle $\dif\Phi_H$ yields a factor $2\pi$ due to the symmetry inside the circularly polarized laser field. We thus obtain the final expression for the total cross section,
\begin{align}\label{totcrosssection_cp}
 \sigma &= \sum_n \int_{E_H^{\text{min}}}^{E_H^{\text{max}}}\dif E_H \left.\frac{\pi\alpha^2M_Z^2t_n}{32\cos^4\theta_W\sin^4\theta_W(q_n^2-M_Z^2)^2}\nonumber\right.\\&\left.\qquad\qquad\times\frac{1}{\sqrt{(q_+q_-)^2-m_*^4}\abs{\vec q_n}}\right.
 \nonumber\\
 &=:\sum_n\sigma_n\,,
\end{align}
which can again be written as a sum over {partial} cross sections $\sigma_n$. The final integration over the produced Higgs boson's energy has been performed numerically.\\

\begin{figure}
  \includegraphics[width=0.45\textwidth]{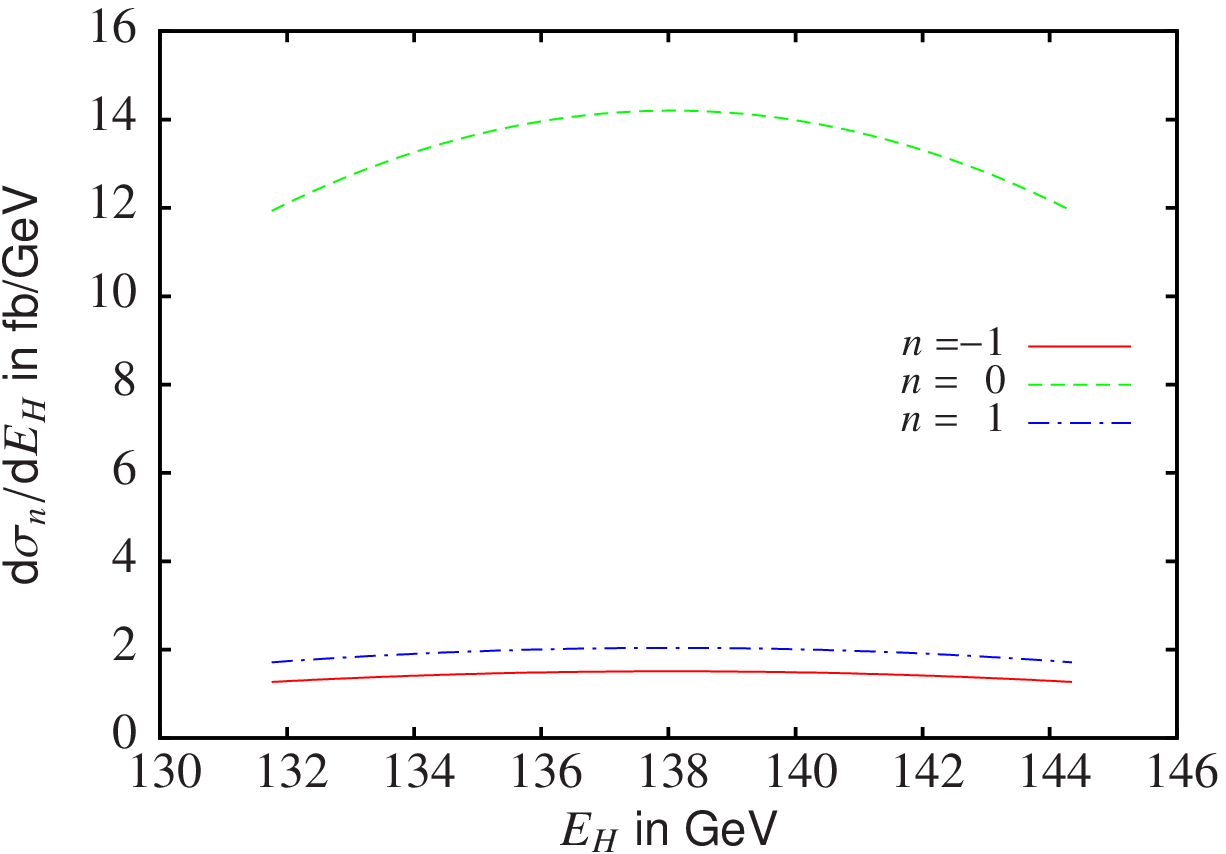}
 \caption{(Color online) Differential cross section as a function of the produced Higgs boson's energy for the three contributing photon orders $n=-1$ (red solid line), $n=0$ (green dashed line) and $n=1$ (blue dashed-dotted line) for $\xi_c=0.5$, $\omega=1$ eV, and $p^0\approx109$ GeV for the process $\ell^+\ell^-\to HZ$ in a circularly polarized laser field. The total cross section is $\sigma\approx212$ fb.}\label{plot_cp_xi0_5_spec}
\end{figure}
\begin{figure}
  \includegraphics[width=0.45\textwidth]{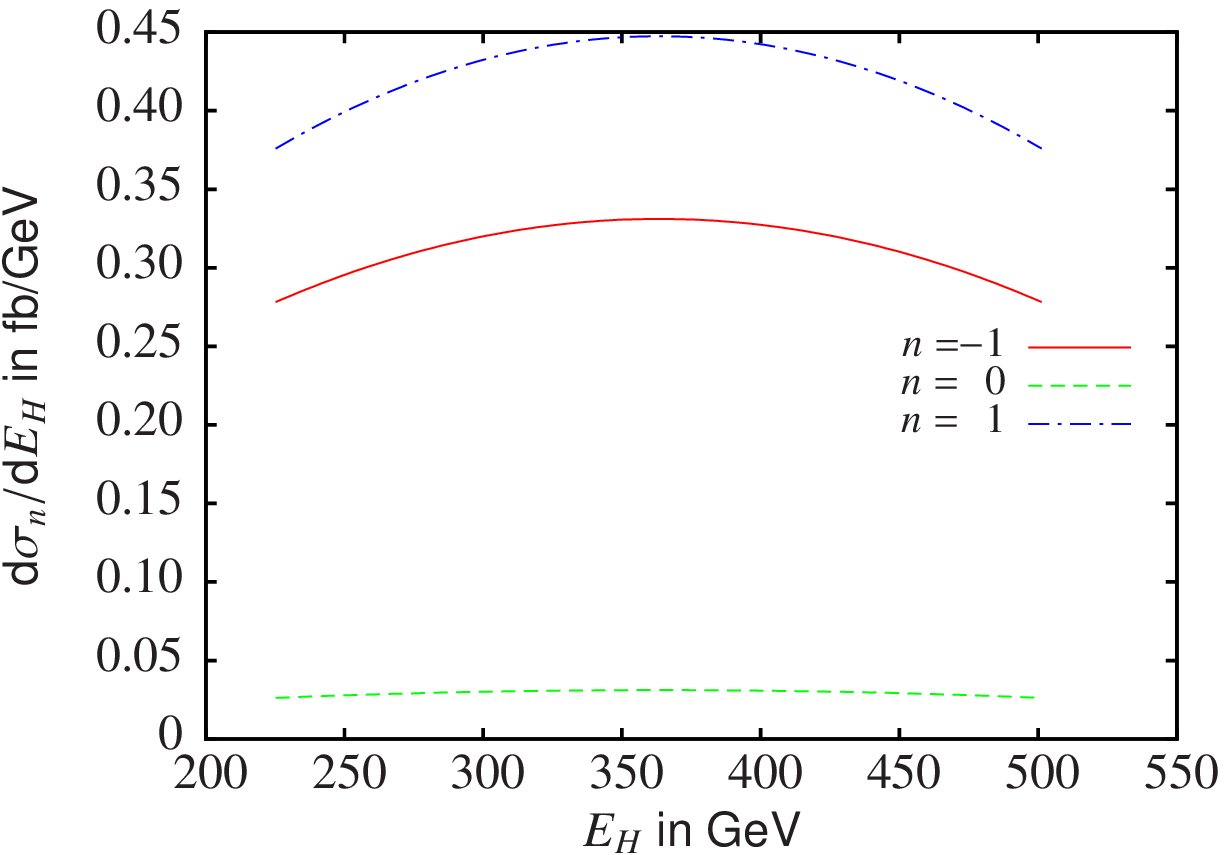}
 \caption{(Color online) Differential cross section as a function of the produced Higgs boson's energy for the three contributing photon orders $n=-1$ (red solid line), $n=0$ (green dashed line) and $n=1$ (blue dashed-dotted line) for $\xi_c=5$, $\omega=1$ eV, and $p^0\approx24$ GeV for the process $\ell^+\ell^-\to HZ$ in a circularly polarized laser field. The total cross section is $\sigma\approx212$ fb.}\label{plot_cp_xi5_spec}
\end{figure}

We now discuss the results of the numerical calculation of Eq. \eqref{totcrosssection_cp}. We will restrict our considerations to the special case where $p_\pm=(p^0,0,0,\pm p^3)$, i.e. where the leptons' initial momenta are opposite and equal and perfectly (anti)parallel to the laser propagation direction. In this case, the laser-dressed lepton energies are $q^0_-\approx p^0(1+\xi_c^2)$ for the particle co-propagating with the laser wave and $q_+^0\approx p^0$ for the counter-propagating particle (see Eq. \ref{q0eq}). Since the lepton momenta contain no components in the directions perpendicular to the laser propagation axis, the argument of the Bessel functions is $\overline{\alpha}=0$ [see Eqs. \eqref{betaDef} and \eqref{BessArg_cp}]. All Bessel functions except $J_0$ have a root at the origin, and therefore, only $n=0$ or $\pm1$ laser photons can be involved in the process [see Eq. \eqref{FourierCoeffElectrons_cp}]. Since the energy of one laser photon is negligible as compared to the total energy involved in the process, the collision energy can then be written as
\begin{equation}\label{cp_E_cm}
 E_{\text{cm}}\approx\sqrt{(q_++q_-)^2}\approx2p^0\sqrt{1+\xi_c^2}\,.
\end{equation}
It is now chosen such that a field-free collision of similar c.m. energy would yield the maximum cross section, i.e. $E_{\text{cm}}\approx244.5$ GeV. As can be seen from Eq. \eqref{cp_E_cm}, the stronger the laser field (i.e. the larger $\xi_c$), the stronger is the enhancement of the initial collision energy. \\

\begin{figure}
 \includegraphics[width=0.45\textwidth]{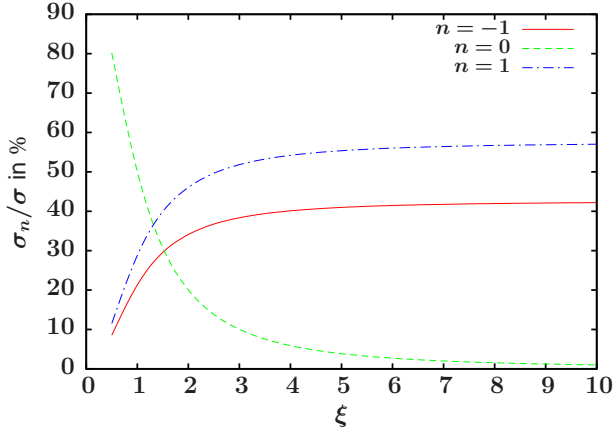}
 \caption{(Color online) Dependence of the contributions from the different photon orders on the laser intensity parameter $\xi_c$ for the process $\ell^+\ell^-\to HZ$ in a circularly polarized laser field where the free lepton energy is set to $p^0=244.5$ GeV$/2\sqrt{1+\xi^2}$, i.e., where the total cross section is the maximum field-free cross section.}\label{plot_cp_xidep}
\end{figure}

Figure \ref{plot_cp_xi0_5_spec} shows the partial cross sections differential in the Higgs boson's energy for each contributing photon order for $\xi_c=0.5$, $\omega=1$ eV, and $p^0\approx109$ GeV, leading to $E_{\text{cm}}\approx244.5$ GeV. When integrated over the Higgs boson's energy and summed over all photon orders, the total cross section is $\sigma\approx212$ fb, i.e. the maximally achievable field-free cross section. Figure \ref{plot_cp_xi5_spec} shows the same for $\xi_c=5$ and $p^0\approx24$ GeV, where the same cross section is obtained. Here, the dominant photon order stems from the absorption of one laser photon, while zero absorption or emission (the dominant order for the smaller laser intensity parameter) plays only a minor role for the stronger laser field. The dependence of the photon orders' contributions to the total cross section on the laser intensity parameter is shown in Fig. \ref{plot_cp_xidep}. The stronger the laser field, the less important is the contribution from $n=0$ and the more important becomes the interaction with the laser field. The total cross section is always the same as for a field-free collision with the total energy $E_{\text{cm}}$ from Eq. \eqref{cp_E_cm}.

\subsection{Linear Polarization}

We now consider the case where the electric field of the laser wave is linearly polarized and the laser potential reads
\begin{equation}\label{laserpot_lp}
 A_l(\kappa):=a_l\cos\kappa\,,
\end{equation}
with $a_l=(0,a,0,0)$. The laser intensity parameter \eqref{xidefallg} for this potential reads
\begin{equation}\label{xidef_lp}
 \xi_l = \frac{ea}{\sqrt{2}m}
\end{equation}
and the Volkov states \eqref{VolkovStateAllgLepton} and \eqref{VolkovStateAllgAntilepton} become
\begin{align}\label{VolkovstatesLinearpol}
 \psi_{\pm}(x)= &\sqrt{\frac{1}{2q^0_\pm}}\left(1\pm\frac{e\dirac k\dirac a_l\cos\kappa}{2(kp_\pm)}\right)u_\pm\nonumber\\
 &\times\e^{\pm\ii(q_\pm x) } \e^{\left(\frac{\ii e(a_lp_\pm)}{(kp_\pm)}\sin\kappa-\frac{e^2a^2}{8(kp_\pm)}\sin(2\kappa)\right)}\,.
 \end{align}
The leptonic current is now
 \begin{align}
  \mathcal{J}_\mu&=\frac{1}{2\sqrt{q_+^0q_-^0}}\nonumber\\
 &\quad\times\int\dif^4x\overline{u}_{+}\left(\Gamma^\mu + \frac{e}{2}\Bigl(\frac{\dirac a_l \dirac k \Gamma^\mu}{(kp_+)} - \frac{\Gamma^\mu\dirac k\dirac a_l}{(kp_-)}\Bigr)\cos\kappa\right. \nonumber\\
&\left.\qquad\qquad- \frac{e^2\dirac a_l\dirac k \Gamma^\mu\dirac k\dirac a_l}{4(kp_+)(kp_-)}\cos^2\kappa\right)u_{-}\nonumber\\
&\quad\times\e^{\left(-\ii\bigl(q_++q_-\bigr)x\right)}\e^{\left(-\ii\bigl(\tilde\alpha_1\sin\kappa+\tilde\alpha_2\sin(2\kappa)\bigr)\right)}\,,
\label{ElectronCurrentExpansion_linpol}
 \end{align}
with the abbreviations
\begin{align}\label{lp_alhpadef}
 \tilde\alpha_1&:=e\Bigl(\frac{(a_lp_+)}{(kp_+)}-\frac{(a_lp_-)}{(kp_-)}\Bigr)\qquad \text{and}\nonumber\\
\tilde\alpha_2&:=\frac{e^2a^2}{8}\Bigl(\frac{1}{(kp_+)}+\frac{1}{(kp_-)}\Bigr)\,.
\end{align}
The functions $\tilde f(\kappa):= \exp(-\ii (\tilde\alpha_1\sin\kappa + \tilde\alpha_2\sin(2\kappa)))$, $\cos\kappa\tilde f(\kappa)$, and $\cos^2\kappa\tilde f(\kappa)$ from Eq. \eqref{ElectronCurrentExpansion_linpol} can again be expanded in a Fourier series,
\begin{align}\label{FourierElectrons_linpol}
 \tilde f(\kappa)&=\sum_{n=-\infty}^\infty \tilde b_n \e^{-\ii n\kappa}\,,\nonumber\\
  \cos(\kappa) \,\tilde f(\kappa)&=\sum_{n=-\infty}^\infty \tilde c_n \e^{-\ii n\kappa}\,,\nonumber\\
  \cos^2(\kappa) \,\tilde f(\kappa)&=\sum_{n=-\infty}^\infty\tilde d_n \e^{-\ii n\kappa}\,
\end{align}
with the coefficients
\begin{align}\label{FourierCoeffElectrons_linpol}
 \tilde b_n&=\tilde J_n(\tilde \alpha_1,\tilde \alpha_2)\,,\nonumber\\
 \tilde c_n&=\frac{1}{2}\biggl(\tilde J_{n-1}(\tilde \alpha_1,\tilde \alpha_2)+\tilde J_{n+1}(\tilde \alpha_1, \tilde\alpha_2)\biggr)\nonumber\\
 \tilde d_n&=\frac{1}{4}\biggl(\tilde J_{n-2}(\tilde\alpha_1, \tilde\alpha_2)+2\tilde J_{n}(\tilde\alpha_1,\tilde \alpha_2) + \tilde J_{n+2}(\tilde\alpha_1, \tilde\alpha_2)\biggr)\,.
\end{align}
Here, the generalized Bessel functions \cite{Reiss}
\begin{equation}\label{genbess_lp}
 \tilde J_n(\tilde\alpha_1, \tilde\alpha_2) := \sum_{\ell=-\infty}^\infty J_{n-2\ell}(\tilde\alpha_1)J_\ell(\tilde\alpha_2)\,,
\end{equation}
which are composed of products of regular cylindrical Bessel functions, occur. Similarly to the calculation for circular polarization, we find 
 
\begin{align}\label{totcrosssection_linpol}
 \sigma &= \sum_n \!\int_{E_H^{\text{min}}}^{E_H^{\text{max}}}\!\!\!\!\dif E_H \left.\frac{\pi\alpha^2M_Z^2{\tilde t}_n}{32\cos^4\theta_W\sin^4\theta_W(q_n^2-M_Z^2)^2}\right.\nonumber\\
 &\left.\qquad\qquad\times\frac{1}{\sqrt{(q_+q_-)^2-m_*^4}\abs{\vec q_n}}\right.\nonumber\\
 &=\sum_n\sigma_n\,
\end{align}
with the same values of $\cos\Theta_H$, $E_H^{\text{min}}$, and $E_H^{\text{max}}$ as for circular polarization [see Eqs. \eqref{higgscos_cp} and \eqref{higgsenergylimits_cp}]. The main difference lies in the trace product $\tilde t_n$ which now contains the expansion from Eq. \eqref{FourierElectrons_linpol}. Eq. \eqref{totcrosssection_linpol} has been evaluated numerically.\\

\begin{figure}
 \includegraphics[width=0.45\textwidth]{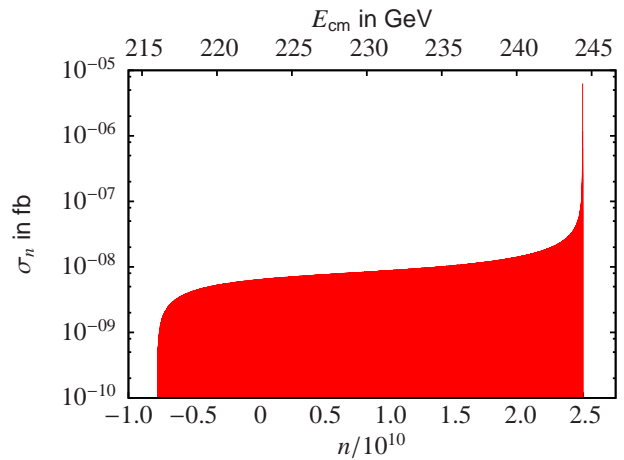}\caption{(Color online) Partial cross sections $\sigma_n$ as function of the photon order $n$ (lower $x$ axis) for $\xi_l=0.5$, $\omega=1$ eV, and $p^0\approx100$ GeV for the process $\ell^+\ell^-\to HZ$ in a linearly polarized laser field. On the upper $x$ axis, the corresponding c.m. energy is shown [see Eq. \eqref{Ecm_n}]. The total cross section amounts to 93 fb.}\label{plot_xi0_5_sigman}
\end{figure}

As before, we now restrict our considerations to the case where the incident leptons' momenta are opposite and equal, $p_\pm=(p^0,0,0,\mp p^3)$, and (anti)parallel to the laser wave vector $\vec k$. In this case, the arguments of the generalized Bessel functions are $\tilde \alpha_1 = 0$ and $\tilde \alpha_2=p^0\xi_l^2/2\omega$. Since all regular Bessel functions except $J_0$ have a root at the origin, the summation in \eqref{genbess_lp} collapses and reduces to $\tilde J_n(0, \tilde\alpha_2)=J_{\ell=n/2}(\tilde\alpha_2)$ with the restriction that $n$ is even (see also Eq. B6 in \cite{Reiss}). This means that in this particular collision geometry, only pairs of laser photons can be absorbed or emitted. Unlike in the case of circular polarization where $\abs{n}\le1$, the total numbers of absorbed or emitted laser photons may in principle be very large. Therefore, the collision energy 
\begin{equation}\label{Ecm_n}
 E_{\text{cm}}(n)=\sqrt{(q_++q_-+nk)^2}
\end{equation}
may vary over a large range of $n$. For large orders, the Bessel functions have a maximum where order and argument are of comparable size (c.f. \cite{Abramowitz}). Accordingly, the numerical calculation of the partial cross sections shows a maximum at $n_{\text{max}}=2\tilde\alpha_2=p^0\xi_l^2/\omega$ and the partial cross sections quickly vanish for higher photon orders (see Fig. \ref{plot_xi0_5_sigman}). The collision energy at this maximum photon order is
\begin{equation}\label{Ecm_nmax}
 E_{\text{cm}}(n_{\text{max}})=2p^0\sqrt{1+2\xi_l^2}\,.
\end{equation}
In principle, the distribution of the partial cross sections has another maximum at $-n_{\text{max}}$, corresponding to $E_{\text{cm}}(-n_{\text{max}})=2p^0$; however, depending on the initial lepton energy, there may occur a cutoff where the collision energy $E_{\text{cm}}(n)$ for a certain photon order does not suffice to exceed the reaction threshold. \\

The striking difference between the distributions of the partial cross sections $\sigma_n$ for linear and circular polarization can be understood by a classical consideration. In classical terms (see, e.g., \cite{dipiazzareview, Itzykson}), the collision energy generally depends on the laser phase $\kappa$ and is given by

\begin{equation}\label{classicalEcm}
 E_{cm}(\kappa) = 2p^0\sqrt{1-\frac{e^2}{m^2}A^2(\kappa)}\,.
\end{equation}
For circular polarization, we have $-e^2A^2(\kappa)=m^2\xi_c^2$ which is constant, in accordance with our finding in Section \ref{HiggsCPsection}. Contrary to that, for linear polarization, we have $-e^2A^2(\kappa)=2m^2\xi_l^2\cos^2(\kappa)$, leading to a minimal collision energy of $2p^0$ for $\cos^(\kappa)=0$ and, for $\cos^2(\kappa)=1$, a maximum collision energy in agreement with Eq. \eqref{Ecm_nmax}. Classically, all collision energies lying in between are allowed, as long as the threshold energy of $(M_Z+M_H)$ is exceeded.\\

\begin{figure}
 \includegraphics[width=0.45\textwidth]{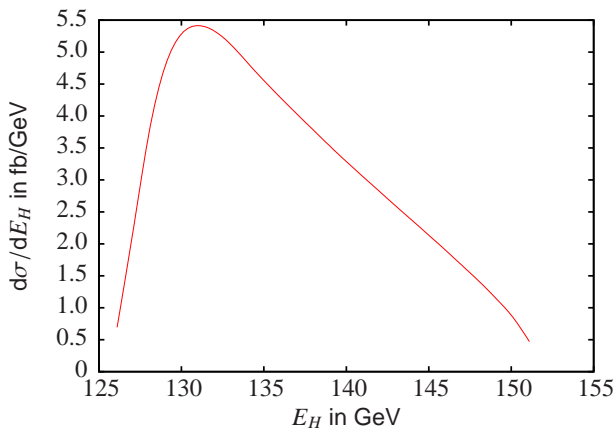}\caption{(Color online) Differential cross section $\dif\sigma/\dif E_H$ as function of the produced Higgs boson's energy $E_H$ for $\xi_l=0.5$, $\omega = 10$ keV, and $p^0\approx100$ GeV for the process $\ell^+\ell^-\to HZ$ in a linearly polarized laser field.}\label{plot_xi0_5_spectrum}
\end{figure}

\begin{figure}
 \includegraphics[width=0.45\textwidth]{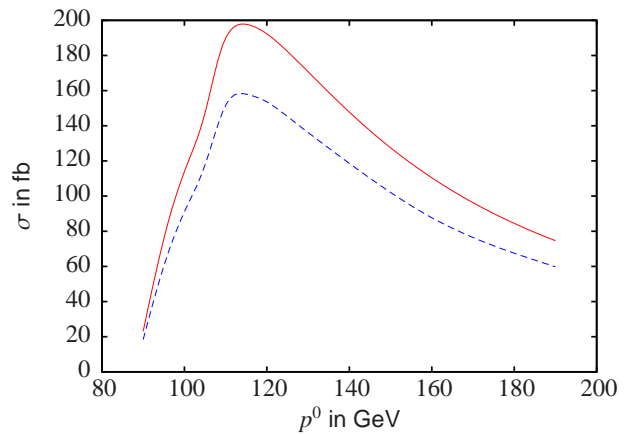}
 \caption{(Color online) Total cross section $\sigma$ as function of the free lepton energy $p^0$ for $\xi_l=0.5$ and $\omega=10$ keV (blue dashed line) for the process $\ell^+\ell^-\to HZ$ in a linearly polarized laser field and corresponding phase average of the field-free cross section according to Eq. \eqref{ff_convolution} (red solid line).}\label{plot_xi0_5_pdep}
\end{figure}

Figure \ref{plot_xi0_5_sigman} shows the partial cross sections' dependence on the photon order $n$ for $\xi_l=0.5$, $\omega=1$ eV, and $p^0\approx100$ GeV. The initial lepton energy is chosen such that $E_{\text{cm}}(n_{\text{max}})\approx244.5$ GeV, i.e. the maximum collision energy would lead to the maximum cross section in the field-free case. Although the distribution looks continuous, it consists of discrete peaks for each even photon number $n$. The corresponding distribution of the produced Higgs boson's energy is shown in Fig. \ref{plot_xi0_5_spectrum} for a photon energy of $\omega=10$ keV which leads to the same total cross section but for which the computation time is vastly reduced. It is asymmetric with a preference toward smaller Higgs boson energies, because only few photon orders contribute to the spectrum for larger Higgs boson energies. {For the same reason, the angular distribution of the produced Higgs boson is narrower, with a tendency towards the laser propagation direction. Furthermore, the electrically charged decay products of the Higgs boson will themselves interact with the laser field, resulting in a so-called ``channeling'' into narrow angular regions \cite{lotstedt}. This effect might result in a smaller area to be covered by detector facilities.}\\
When summed over all photon orders, the total cross section for the parameters from Fig. \ref{plot_xi0_5_sigman} amounts to 93 fb, which is significantly smaller than the maximum field-free cross section. This discrepancy can be understood by investigating the dependence of the total cross section on the free lepton energy $p^0$ as shown in Fig. \ref{plot_xi0_5_pdep}. The shape of the curve in Fig \ref{plot_xi0_5_pdep} can be understood qualitatively as follows: The minimum collision energy inside the laser field amounts to $E_\text{cm}(-n_{\text{max}}) \approx 2p^0$ [see also the classical energy in Eq. \eqref{classicalEcm}]. When it lies below the threshold energy for $HZ$ production, the partial cross sections associated with the lower range of photon numbers $(n\gtrsim -n_\text{max})$ cannot contribute, this way reducing the total cross section. This situation is also illustrated in Fig. \ref{plot_xi0_5_sigman}. When $p^0$ is increased, more and more $n$ values are kinematically allowed, leading to a growth of the total cross section. The maximum cross section in Fig. \ref{plot_xi0_5_pdep} is reached when $2p^0$ is close to 244.5 GeV. Here, all even $n\in [-n_\text{max} , n_\text{max}]$ give contributions to the total cross section, with the corresponding collision energies lying in the region where the field-free cross section in Fig. \ref{ff_sigplot} peaks. Accordingly, a total cross section close to (but below) the optimum value $\sigma_\text{ff} \approx 212$ fb is reached. When we increase $p^0$ further, still all even $n\in [-n_\text{max} , n_\text{max}]$ contribute but we move towards higher collision energies and, thus, slide down the cross section curve in Fig. \ref{ff_sigplot}. This explanation is backed by the comparison with a phase average of the field-free cross section, depicted in Fig. \ref{plot_xi0_5_pdep}, 
\begin{equation}\label{ff_convolution}
 \sigma_\text{ave} = \frac{1}{2\pi}\int_0^{2\pi}\sigma_\text{ff}(E_\text{cm}(\kappa))\dif\kappa
\end{equation}
with the field-free cross section $\sigma_{\text{ff}}$ as introduced in Eq. \eqref{sigma_ff} and the phase-dependent collision energy as defined in Eq. \eqref{classicalEcm}. In the integral, only the phases $\kappa$ for which $E_{\text{cm}}(\kappa)>(M_Z+M_H)$ are counted. The qualitative run of the curve as well as the height of the maximum resemble the ones found in the laser-dressed calculation. \\ 
\begin{figure}
 \includegraphics[width=0.45\textwidth]{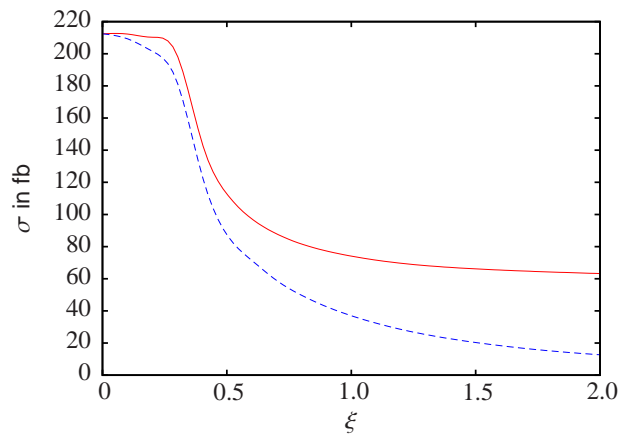}
 \caption{(Color online) Total cross section as a function of the laser intensity parameter for $\omega=10$ keV and $p^0=244.5$ GeV$/2\sqrt{1+2\xi_l^2}$ (blue dashed line) for the process $\ell^+\ell^-\to HZ$ in a linearly polarized laser field and corresponding phase average of the field-free cross section according to Eq. \eqref{ff_convolution} (red solid line). }\label{plot_xidep}
\end{figure}

Increasing the laser intensity parameter, the initial lepton energy can be further reduced. The total cross section as a function of the laser intensity parameter $\xi_l$ is shown in Fig. \ref{plot_xidep}. For each $\xi_l$, the free lepton energy $p^0$ is chosen such that the maximum collision energy \eqref{Ecm_nmax} is equal to 244.5 GeV which would correspond to the maximum field-free cross section. Also shown in Fig. \ref{plot_xidep} is the corresponding phase average of the field-free cross section as introduced in Eq. \ref{ff_convolution}. We find that, for small laser intensity parameters $\xi_l$, the two curves are very similar.  For $\xi_l\gtrsim 0.5$, both curves begin to deviate from each other. While the cross section prediction from our full quantum calculation continues to decrease, the phase-averaged cross section reaches saturation for large values of $\xi_l$. The latter can be understood as follows. With our choice of $p^0$, in order to exceed the reaction threshold, the phase $\kappa$ must fulfill
\begin{equation}
 \cos^2\kappa > \frac{(1+2\xi_l^2)R^2-1}{2\xi_l^2}
\end{equation}
with $R=244.5$ GeV$/216$ GeV. For large $\xi_l$, this condition becomes $\cos^2\kappa > R^2$ and thus the contributing phase interval is constant, leading to a saturation of the field-free phase average.

\subsection{Laser parameters}\label{HiggsParamsSec}

\begin{table*}
\centering
\small
\begin{tabular}{rcccccccc}
\hline
 Polarization & circ. & circ. & lin. & lin. & lin. & lin. & lin. & lin. \\ \hline
Colliding particles&  $e^+e^-$ & $\mu^+\mu^-$ & $e^+e^-$& $\mu^+\mu^-$& $\mu^+\mu^-$ &  $e^+e^-$ & $\mu^+\mu^-$ & $\mu^+\mu^-$\\ \hline
Intensity parameter $\xi$ & 0.5 & 0.5 &  0.5 & 0.5 & 0.5 & 5 & 5 & 5 \\ \hline
Photon energy $\omega$ (eV) & 1 & 1 & 1 & 1 & 10 & 1 & 1 & 10 \\ \hline
Laser intensity $I$ (W/cm$^2$) & $4.5\times10^{17}$ & $1.9\times10^{22}$  & $4.5\times10^{17}$  & $1.9\times10^{22}$ &$1.9\times10^{24}$ & $4.5\times10^{19}$ & $1.9\times10^{24}$ &  $1.9\times10^{26}$\\ \hline
Free lepton energy $p^0_\pm$ (GeV) & 109  & 109 & 100&100& 100& 17&17 & 17\\ \hline
Lorentz factor $\gamma$ &  $2\times10^5$  & 1035 & $2\times10^5$ & 945 &945 & $3\times10^4$ & 162 & 162\\ \hline
Beam waist $w_0$ (mm) &   167  &0.81& 153& 0.74 & 0.074 &208& 1.01 & 0.1 \\ \hline
Pulse duration (ns) &  $4.7\times10^{5}$ & 11.1 &$3.9\times10^5$&9.2 & 0.9 &$7.3\times10^{5}$&17.1 & 1.7 \\ \hline
Pulse power (W) & $3.96\times10^{20}$  & $3.92\times10^{20}$  & $2.68\times10^{15}$ & $5.47\times10^{17}$ &$5.47\times10^{17}$ & $3.64\times10^{17}$ &$7.44\times10^{19}$ & $7.44\times10^{19}$ \\ \hline
Pulse energy (J) & $1.87\times10^{17}$ & $4.33\times10^{12}$  &$1.05\times10^{12}$ &$5.05\times10^{9}$ & $5.05\times10^{8}$ & $2.65\times10^{14}$& $1.27\times10^{12}$ & $1.27\times10^{11}$ \\ \hline
\end{tabular}
%
\caption{Minimally required laser parameters for the process $\ell^+\ell^-\to HZ^0$ for the parameters used above with the photon energy $\omega=1$ eV, corresponding to $\lambda=1.24$ $\mu$m. For comparison, we also list the required parameters for $\omega=10$ eV for muon-antimuon collisions in a linearly polarized laser field. }\label{params_1eV}
\end{table*}

We have seen in the previous sections that employing an additional laser field in elementary particle collisions may reduce the required initial lepton energy if large laser intensity parameters are involved. We will now discuss the laser parameters required for such a collider scheme. \\
In our calculation of the Volkov states, an infinitely long plane laser wave was assumed. This approximation is justified only if the leptons sojourn inside the laser field for at least one cycle of the field oscillation. In particular, the laser pulse must cover the co-propagating lepton's trajectory for at least this period. Therefore, the pulse must be long enough so that the distance the lepton travels in the laser propagation direction is covered by the pulse, and the focal area must be large enough so that the leptons are not kicked out of the laser pulse by the interaction with the electric field. The distance in the laser field propagation direction $z$ that the lepton with Lorentz factor $\gamma$ travels during one full cycle of a laser with wavelength $\lambda$ is given by \cite{Chang}
\begin{equation}\label{deltaz}
 \Delta z = 2\gamma^2\lambda(1+\xi^2)
\end{equation}
and its oscillations along the polarization direction have the amplitude
\begin{equation}\label{deltax}
 \Delta x \approx \gamma\xi\lambda\,.
\end{equation}
{It is necessary for the leptons not only to be covered by the laser wave, but to also travel inside a region of high intensity. We therefore consider a Gaussian beam with beam waist $w_0$ and demand that the lepton trajectory be covered by the Rayleigh length $z_R=\pi w_0^2/\lambda$ \cite{Gaussbeam}. We find}
that the beam waist must exceed 
\begin{equation}\label{w0min}
 w_0\ge \text{max}\left\{ \gamma\xi\lambda\,,\,\,\gamma\lambda\sqrt{(1+\xi^2)/\pi}\right\}\,.
\end{equation}
From this, the minimum focal area and pulse duration of the laser pulse can be calculated as shown in Table \ref{params_1eV}. The previous calculations are independent of the lepton mass and therefore hold for both muon-antimuon and electron-positron collisions. For comparison, we list the laser parameters required for both lepton species for the collision parameters from the examples shown above where a photon energy of $\omega=1$ eV and thus a laser wavelength of $\lambda=1.24$ $\mu$m has been assumed. The first two columns in Table \ref{params_1eV} show the parameters for the circularly polarized laser field with an intensity parameter $\xi_c=0.5$. As can be seen, the electron-positron collision requires a smaller laser intensity for the same laser intensity parameter $\xi_c$. However, due to the much larger Lorentz factor, the required laser pulse energy is much larger than for muon-antimuon collisions; it attains an enormous value of about 200 PJ which is clearly far beyond technical capabilities in the foreseeable future. For muons, it lies in the TJ regime, which is much smaller but still out of reach for near-future laser facilities. The right-hand side of Table \ref{params_1eV} show the same for our examples in the linearly polarized laser field. Here, an {\em elliptic Gaussian beam} according to section 6.12 in \cite{yariv} is assumed where the beam extension along the polarization direction of the electric field is assumed to be $w_0$ according to Eq. \eqref{w0min} and perpendicular to it, it is assumed to be $\lambda$ \cite{footnotebeam}. This leads to a much smaller focal area and, thus, to vastly reduced laser pulse powers $P=\pi w_0\lambda I$ and energies as compared to the circularly polarized field where $P=\pi w_0^2I$. Therefore, while the biggest advantage of the circularly polarized laser field lies in its reproduction of the field-free cross section and thus possibly high number of produced particles, from a technical point of view the considered collider scheme is more practicable with a linearly polarized laser field. 
\\
As can be seen in Table \ref{params_1eV}, the profit of the superimposed laser field may be larger in muon collisions than electron collisions since, due to the smaller Lorentz factor, smaller laser pulse durations and thus pulse powers are involved here. In particular, the required pulse energies are substantially reduced to $\sim 1$ GJ here. Additionally, damping effects such as Compton scattering off the laser beam are less prominent for the much heavier muons and antimuons.\\
We note that increasing the photon energy, while increasing the required laser intensity for a fixed laser intensity parameter $\xi$, leads to a smaller required pulse energy: while the intensity for a fixed intensity parameter scales with $I\propto\omega^2$, both the wavelength (i.e. beam waist in $y$ direction) and the required beam waist in $x-$direction scale with its inverse. Thus, the pulse power remains independent of the photon energy. However, the pulse duration is again inversely proportional to $\omega$, and therefore, the pulse energy can be reduced. The effect is shown in columns 5 and 8 of Table \ref{params_1eV} where, for muons, the required parameters for $\omega=10$ eV are listed. \\

Another positive effect of a higher photon energy lies in the fact that the focal area of the laser pulse and thus the interaction area scales with the square of the laser wavelength. A large cross sectional area not only requires high-power laser pulses, but also reduces the luminosity of the collider, since the latter depends on the mean distance between the colliding particles and thus the diameter of the particle beam. For example, in the case of circular polarization and for the parameters of the second column in Table \ref{params_1eV}, the reduction of the collider luminosity due to this effect is approximately two orders of magnitude, if we consider a muon beam with 200 $\mu$m radius as is realistic for the considered initial energy \cite{alsharoa}. 
If we consider, e.g., the parameters from column 7 in Table \ref{params_1eV}, i.e. a muon-antimuon collision in a linearly polarized laser beam with the intensity parameter $\xi_l=5$, the beam waist in $x-$direction is larger than the initial muon beam, resulting in a luminosity loss of approximately one order of magnitude. In the perpendicular direction, the beam waist $\lambda$ is much smaller than the initial muon beam. This leads to a reduced luminosity due to the fact that not all the muons from the initial beam will interact with the laser pulse and therefore the number of colliding particles with sufficient energy is decreased. Increasing the photon energy and thus reducing the minimally required beam waist may be beneficial for the luminosity loss; if the available pulse power and energy become large enough, a focal area of the order of the initial muon beam size might become possible, which could vastly reduce the luminosity loss.

\section{Other Laser-Boosted Processes}\label{otherprocesses}
\begin{figure}
 \centering
 \subfloat[]{\includegraphics[width=0.23\textwidth]{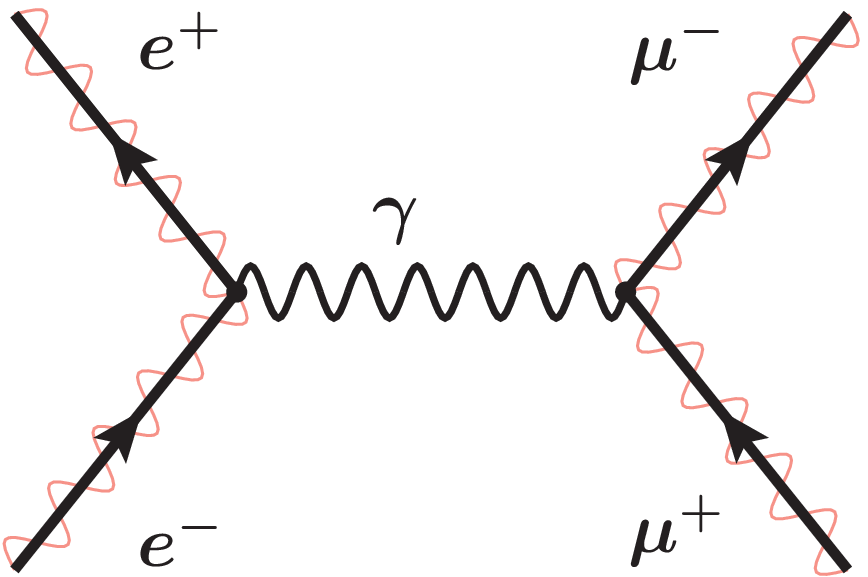}}
 \hfill
 \subfloat[]{\includegraphics[width=0.23\textwidth]{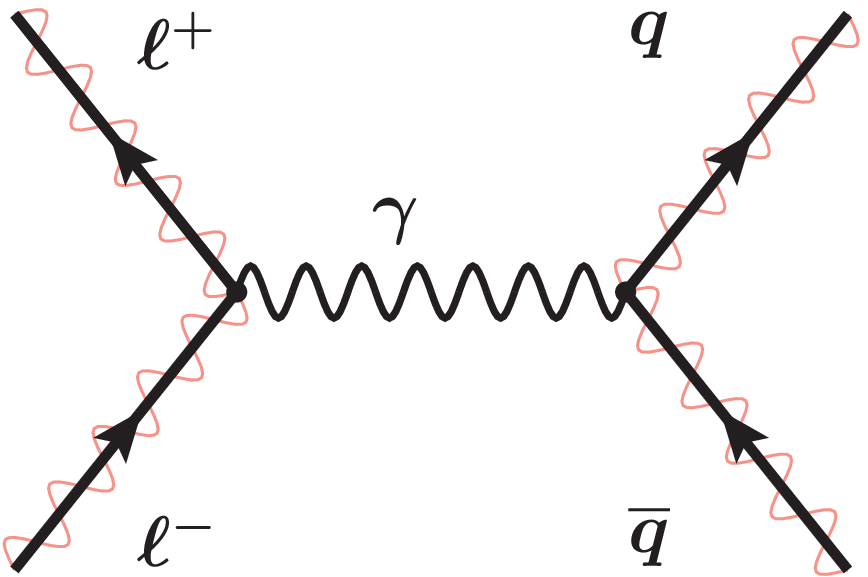}}
 \caption{(Color online) Feynman diagrams for the processes $e^+e^-\to \mu^+\mu^-$ (a) and $\ell^+\ell^-\to \text{hadrons}$ (b). The wiggled lines indicate that the (electrically charged) leptons interact with a superimposed laser field. }\label{OP_Feyns}
\end{figure}

We have seen in the last section that, while the combination of conventional accelerator techniques with strong laser pulses may offer interesting opportunities, the laser intensities and pulse energies required for the production of Higgs bosons are very challenging for near-future experimental implementation. Therefore, in order to test the validity of the underlying physical principles, the study of other, more easily realizable processes might be of interest. In this section, we will therefore first address the production of muon-antimuon pairs in laser-boosted electron-positron collisions in detail and then discuss the possibility of hadron production at the future ELI (see Fig. \ref{OP_Feyns}).
\subsection{Muon pair production}

In principle, the calculation of the process $e^+e^-\to\mu^+\mu^-$ inside a laser field is similar to the one for Higgs boson production presented above. The main differences are the virtual particle propagating between the initial and outgoing particles' vertices and the fact that the produced particles carry electric charge and thus interact with the laser field. The former is, in this case, a photon instead of a $Z$ boson which is represented in the transition amplitude by a free photon propagator. The latter is taken account for by employing Volkov states for the description of the produced muon-antimuon pair, allowing for the absorption of laser photons at their vertex as well. Please note, however, that since the muons are much heavier than the electrons (muon mass $M\approx200m$), they are influenced by the laser field only with an intensity parameter $\Xi=(m/M)\xi\approx\xi/200$.\\

The transition amplitude then reads 
\begin{align}
  \mathcal{S} = -\ii\alpha\iint & \overline{\psi}_{+}(x)\gamma^\mu\psi_{-}(x)\\ \nonumber
  &\mathcal{D}_{\mu\nu}(x-y)\overline{\Psi}_{-}(y)\gamma^\nu\Psi_{+}(y)\dif^4x\,\dif^4y\,
\end{align}
with the free photon propagator
\begin{equation}\label{photonpropagator}
 \mathcal D_{\mu\nu}(x-y)=\lim_{\epsilon\to+0}\int \frac{\dif^4q}{(2\pi)^4} \frac{4\pi\e^{\ii q(x-y)}g^{\mu\nu}}{q^2+\ii \epsilon}\,.
\end{equation}
The Volkov states $\psi_\pm$ for the electron and positron as well as $\Psi_\pm$ for the produced muon-antimuon pair depend on the polarization of the laser field and the transition amplitude is evaluated again for circular and linear polarization.

\subsubsection{Circular polarization}

Like in the case of the Higgs boson production, the current of the incoming leptons is expanded in a Fourier series. For circular polarization, the same result for $^{e^\pm\!\!\!}\mathcal{J}_\mu$ is obtained as in Eq. \eqref{FourierElectrons_cp}, only the matrices $\Gamma^\mu$ have to be replaced by the Dirac matrices $\gamma^\mu$. The muon-antimuon current is now treated in a similar way. It reads
\begin{align}
 ^{\mu^\pm\!\!\!}{\mathcal{J}}^\nu&=\frac{1}{2\sqrt{Q_+^0Q_-^0}}\nonumber\\
 &\quad\int\dif^4y\overline{U}_{-}\left[\Bigl(\gamma^\nu-\frac{e^2a^2k^\nu}{2(kP_+)(kP_-)}\dirac k\Bigr)\right. \nonumber\\
&\qquad+\frac{e}{2} \Bigl(\bigl(\frac{\gamma^\nu\dirac k\dirac a_1}{(kP_+)} - \frac{\dirac a_1\dirac k\gamma^\nu}{(kP_-)}\bigr)\cos\eta \nonumber\Bigr.\\
&\qquad\,\left.\Bigl. + \bigl(\frac{\gamma^\nu\dirac k \dirac a_2}{(kP_+)}-\frac{\dirac a_2\dirac k\gamma^\nu}{(kP_-)}\bigr)\sin\eta\Bigr)\right]U_{+}\nonumber\\
&\quad\times \e^{\ii(Q_++Q_-)y}\times\e^{-\ii(\beta_1\sin\eta - \beta_2\cos\eta)}
\label{MuonsCurrentExpansion_cp}
\end{align}
with $\eta:=(ky)$, the muon/antimuon free Dirac spinors $U_\pm$, their free (dressed) momenta $P_\pm$ ($Q_\pm$), and
\begin{align}
 \beta_j = \frac{(a_jP_-)}{(kP_-)}-\frac{(a_jP_+)}{(kP_+)}\,.
\end{align}
Note that the sign differs from the corresponding definition \eqref{betaDef}. The periodic functions in \eqref{MuonsCurrentExpansion_cp} can be expanded into Fourier series analogously to Eqs. \eqref{FourierElectrons_cp}--\eqref{kappa_0_cp}. The photon index at the muon-antimuon vertex will be denoted as $N$ and the Bessel functions' argument $\overline{\beta}=\sqrt{\beta_1^2+\beta_2^2}$.\\
The muonic current can then be written as 
\begin{equation}
  ^{\mu^\pm\!\!\!}\mathcal{J}^\nu=\frac{1}{2\sqrt{Q_+^0Q_-^0}} \sum_{N=-\infty}^\infty \int\dif^4y\,\,{^{\mu^\pm\!\!\!}\mathcal{M}}_N^\nu\cdot\e^{\ii(Q_++Q_--Nk)y}\,
\end{equation}
with the spinor-matrix product ${^{\mu^\pm\!\!\!}\mathcal{M}}_N^\nu$, and the transition amplitude becomes
\begin{align}
 \mathcal S =\frac{-\ii\alpha(2\pi)^5}{2\sqrt{q_+^0q_-^0Q_+^0Q_-^0}}\sum_{r,n}&{^{\mu^\pm\!\!\!}\mathcal{M}^\nu_{r-n}}{^{e^\pm\!\!\!}\mathcal{M}_\nu^n}\nonumber\\
 &\cdot\frac{\delta(q_++q_-+rk-Q_+-Q_-)}{(q_++q_-+nk)^2}\,.
\end{align}
Here, $r=n+N$ is the total number of absorbed laser photons, consisting of the number of absorbed photons at the electronic and muonic vertices, $n$ and $N$, respectively. The energy of the virtual photon propagating between the two vertices is given by $(q_++q_-+nk)=(Q_++Q_--Nk)$ and the total four-momentum transferred to the produced muons is 
\begin{equation}
 q_r=q_++q_-+rk\,.
\end{equation}
The derivation of the partial cross sections $\sigma_r$ is very similar to the procedure for Higgs boson creation and leads to 
\begin{align}\label{mu_cp_sigr}
 \sigma_r &=\frac{1}{8\sqrt{(q_+q_-)^2-m_*^4}} \nonumber\\
 &\cdot\int_{E_-^{\text{min}}}^ {E_-^{\text{max}}}\!\!\!\!\dif Q_-^0\sum_{nn'}\frac{\pi\alpha^2T^r_{nn'}}{(q_++q_-+nk)^2(q_++q_-+n'k)^2\abs{\vec q_r}}
\end{align}
with the trace product 
\begin{equation}
 T^r_{nn'}:= \sum_{\text{spins}}({^{e^\pm\!\!\!}\mathcal{M}_\rho^{n'}})^\dagger({^{\mu^\pm\!\!\!}\mathcal{M}^\rho_{r-n'}})^\dagger{^{\mu^\pm\!\!\!}\mathcal{M}^\nu_{r-n}}{^{e^\pm\!\!\!}\mathcal{M}_\nu^n}\,,
\end{equation}
 the produced muon's emission angle given by
\begin{equation}\label{mu_cp_cosTheta}
 \cos\Theta_{-}^0=\frac{2q_r^0Q_-^0-(q_r)^2}{2\abs{\vec q_r}\abs{\vec Q_-}}\,,
\end{equation}
and the integration limits
\begin{align}
 E_-^{\text{min}}&:=\frac{q_r^0}{2}-\frac{\abs{\vec q_r}}{2}\sqrt{1-\frac{4M_*^2}{(q_r)^2}}\quad\text{and}\nonumber\\
 E_-^{\text{max}}&:=\frac{q_r^0}{2}+\frac{\abs{\vec q_r}}{2}\sqrt{1-\frac{4M_*^2}{(q_r)^2}}\,
\end{align}
with the muons' effective mass $M_*=\sqrt{1+\Xi^2}M$ in the laser field.

For the numerical evaluation of Eq. \eqref{mu_cp_sigr}, we set again $p_+^3=-p_-^3$, leading to $\overline{\alpha}=0$ and thus $n,n'\in\{-1,0,1\}$. In contrast to the production of the electrically neutral bosons discussed above, there now is the possibility of absorption or emission of laser photons at the produced muons' vertex. The number of these photons must then be $N,N'\in\{r-1,r,r+1\}$, and there is no principle restriction of the total number $r$ of absorbed or emitted laser photons.\\
The collision energy as a function of the total number of absorbed or emitted photons is given by
\begin{equation}\label{muons_Ecm_r}
 E_{\text{cm}}(r) = \sqrt{(q_++q_-+rk)^2}\,.
\end{equation}
Like in the Higgs boson production process considered above, the total cross section in a circularly polarized laser field is always given by the field-free cross section obtained for the collision energy for zero absorption,
\begin{equation}
 E_{\text{cm}}(r=0)=\sqrt{(q_++q_-)^2}\approx2p^0\sqrt{1+\xi_c^2}\,.
\end{equation}
The field-free cross section \cite{PeskinSchroeder} for the considered process $e^+e^-\to\mu^+\mu^-$ has a maximum of $\sigma_{\text{ff}}\approx1$ $\mu$b at a collision energy of $E_{\text{cm}}\approx250$ MeV. Without the absorption or emission of any laser photons but simply by employing the laser-dressed electron-positron four-momenta $q^\pm$, this collision energy would be obtained for a free energy of $p^0=p^0_\pm\approx88$ MeV for a laser intensity parameter of $\xi_c=1$. \\
\begin{figure}
 \includegraphics[width=0.45\textwidth]{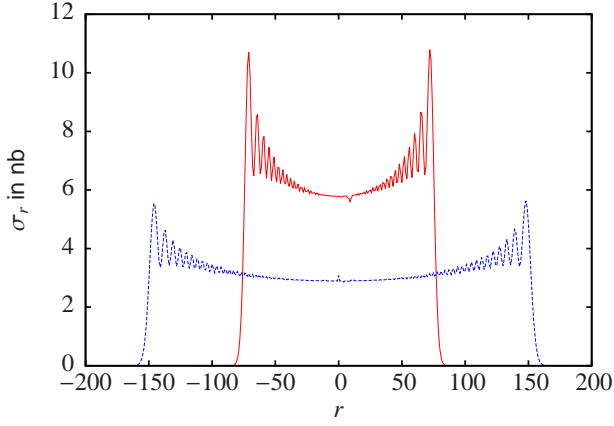}\caption{(Color online) Partial cross sections $\sigma_r$ as a function of the total number of absorbed or emitted laser photons $r$ for $\xi_c=1$ and $p^0\approx88$ MeV for the process $e^+e^-\to \mu^+\mu^-$ in a circularly polarized laser field. The photon energy is set to $\omega = 10$ keV (red solid line) and $\omega=5$ keV (blue dashed line). The total cross section in both cases is $\sigma = \sum_r\sigma_r\approx1$ $\mu$b. }\label{plot_mu_cp_sigr}
\end{figure}

Figure \ref{plot_mu_cp_sigr} shows the partial cross sections $\sigma_r$ as function of the total number of absorbed or emitted laser photons $r$ for $\xi_c=1$ and $p^0\approx88$ MeV for different values of the photon energy $\omega$. It can be seen that a smaller photon energy leads to an increased number of contributing photon orders (since $\bar\beta\propto1/\omega$), while at the same time the absolute value of the partial cross sections is reduced, so that the total cross section $\sigma=\sum_r\sigma_r$ is independent of $\omega$.

\subsubsection{Linear polarization}

\begin{figure}
  \includegraphics[width=0.45\textwidth]{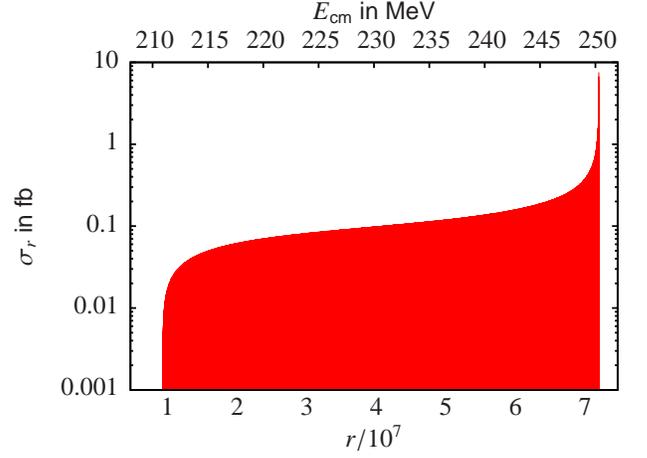}\caption{(Color online) Partial cross sections $\sigma_r$ as a function of the total number of absorbed or emitted laser photons $r$ (lower $x$ axis) for $\xi_l=1$, $\omega=1$ eV and $p^0\approx72.3$ MeV for the process $e^+e^-\to \mu^+\mu^-$ in a linearly polarized laser field. On the upper $x$ axis, the corresponding c.m. energy is shown [see Eq. \eqref{muons_Ecm_r}]. The total cross section is $\sigma = \sum_r\sigma_r\approx210$ nb. }\label{plot_mu_lp_sigr}
\end{figure}

We now consider muon-antimuon pair creation from electron-positron collisions in a linearly polarized laser field. Like before, the current of the incoming particles can be obtained from the similar current \eqref{ElectronCurrentExpansion_linpol} by replacing $\Gamma^\mu$ with $\gamma^\mu$. And, like in the case of circular polarization, the muonic current is obtained in a very similar way,
\begin{align}\label{muons_mucurrent_lp}
 ^{\mu^\pm\!\!\!}\mathcal{J}^\nu=&\frac{1}{2\sqrt{Q^0_+Q_-^0}} \nonumber\\
&\cdot\int\dif^4y\overline{U}_{-}\left(\gamma^\nu + \frac{e}{2}\Bigl(\frac{\gamma^\nu \dirac k \dirac a_l}{(kP_+)}- \frac{\dirac a_l\dirac k\gamma^\nu}{(kP_-)}\Bigr)\cos\eta  \right.\nonumber\\
&\qquad\qquad\qquad\left. - \frac{e^2\dirac a_l\dirac k \gamma^\nu\dirac k\dirac a_l}{4(kP_+)(kP_-)}\cos^2\eta\right)U_{+}\nonumber\\
&\quad\cdot\e^{\ii\bigl(Q_++Q_-\bigr)y}\e^{-\ii\bigl(\tilde\beta_1\sin\eta+\tilde\beta_2\sin(2\eta)\bigr)}\,,
\end{align}
with 
\begin{align}
 \tilde\beta_1 &= e\left(\frac{(a_lP_-)}{(kP_-)} - \frac{(a_lP_+)}{(kP_+)}\right)\nonumber\\
 \tilde\beta_2 &= -\frac{e^2a^2}{8}\left(\frac{1}{(kP_+)}+\frac{1}{(kP_-)}\right)\,.
\end{align}
The Fourier expansion of the periodic functions in the muonic current \eqref{muons_mucurrent_lp} is again performed analogously to Eqs. \eqref{ElectronCurrentExpansion_linpol}-\eqref{genbess_lp}, again containing generalized Bessel functions $\tilde J_N(\tilde\beta_1, \tilde\beta_2)$.\\

\begin{figure}
 \includegraphics[width=0.45\textwidth]{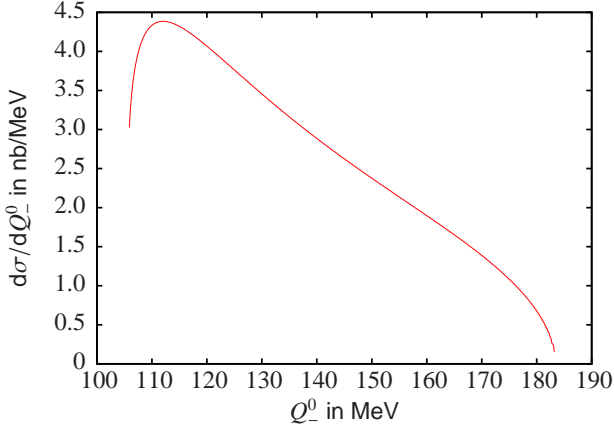}\caption{(Color online) Differential cross section $\dif\sigma/\dif Q_-^0$ as a function of the produced muon's energy $Q_-^0$ for $\xi_l=1$, $\omega = 10$ keV, and $p^0\approx72.3$ MeV for the process $e^+e^-\to \mu^+\mu^-$ in a linearly polarized laser field.}\label{plot_mu_lp_spec}
\end{figure}

The derivation of the cross section, the produced muon's emission angle and the limits for the final integration over the produced muon's energy is the same as in the case for circular polarization, the only difference lying in the spinor-matrix products $^{\mu^\pm\!\!\!}_l\mathcal{M}_N^\nu$ and $^{e^\pm\!\!\!}_l\mathcal{M}_\mu^n$ and thus the trace product $_lT^r_{nn'}$.\\

Again, we performed the numerical calculation of the partial cross sections in a setup where $p^3_+=-p^3_-$. The collision energy is given by $E_{\text{cm}}=\sqrt{q_r^2}$. Again, only pairs of photons can be absorbed or emitted at the electron-positron vertex. However, there are no further constraints on the numbers $n,n'$ interacting at this vertex. Therefore, since $\Xi_l=m/M\xi_l\approx\xi_l/200$, the number of absorbed or emitted laser photons at the muonic vertex is expected to be much smaller than the number of those absorbed or emitted at the electronic vertex. For $\xi_l=1$, photon absorption or emission at the muon-antimuon vertex indeed shows to be negligible, and thus $N=N'=0$. From this follows $n=n'=r$ and there is no double summation in the expression for the partial cross section $\sigma_r$. It can therefore be written as

\begin{align}
 \sigma_r \approx \frac{\pi\alpha^2}{8\sqrt{(q_+q_-)^2-m_*^4}}\!\!\iint\!\!\frac{_lT^r_{rr}}{(q_++q_-+rk)^4\abs{\vec q_r}}\dif Q_-^0\dif\Phi_-.
\end{align}

\begin{figure}
 \includegraphics[width=0.45\textwidth]{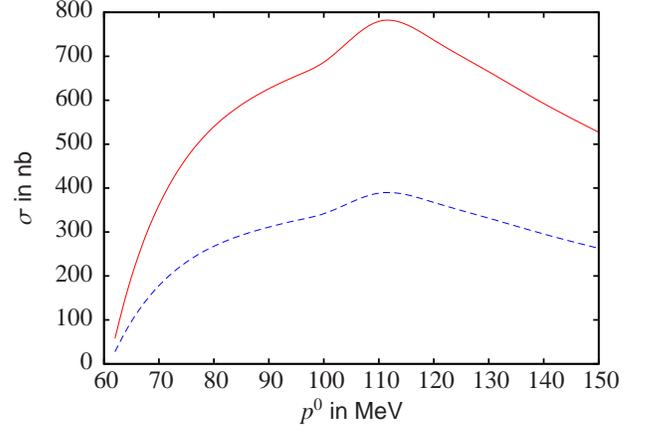}\caption{(Color online) Total cross section $\sigma$ as function of the free lepton energy $p^0$ for $\xi_l=1$ and $\omega=10$ keV (blue dashed line) for the process $e^+e^-\to \mu^+\mu^-$ in a linearly polarized laser field and corresponding field-free phase average (red solid line). }\label{plot_mu_lp_pdep}
\end{figure}

Here, unlike in the Higgs boson production case, the integration over the produced muon's azimuth angle $\Phi_-$ has to be performed explicitly since the muon interacts with the laser field and thus may be influenced by the nonsymmetrical field configuration.\\

The only relevant contribution of the muonic current to the trace product $_lT^r_{rr}$ stems from the term containing $\tilde J_N(\tilde\beta_1,\tilde\beta_2)$ since the other summands are of order $\Xi_l$ or $\Xi_l^2$. Like in the Higgs boson production process, the first Bessel argument in the electronic current vanishes, $\tilde\alpha_1=0$, and the second argument can be written as $\tilde \alpha_2=\xi_l^2p^0/2\omega$. The generalized Bessel function collapses to $\tilde J_r(0,\tilde\alpha_2)=J_{r/2}(\tilde\alpha_2)$, which has a maximum for $r_{\text{max}}/2=\tilde\alpha_2$. The collision energy corresponding to this number of absorbed laser photons, $E_{\text{cm}}(r_{\text{max}})=2p^0\sqrt{1+2\xi_l^2}$, is now again set to obtain the maximum field-free cross section, $E_{\text{cm}}(r_{\text{max}})\approx250$ MeV. For $\xi_l=1$, this is the case for $p^0\approx72.3$ MeV. Figure \ref{plot_mu_lp_sigr} shows the corresponding distribution of the partial cross sections.\\

\begin{figure}
 \includegraphics[width=0.45\textwidth]{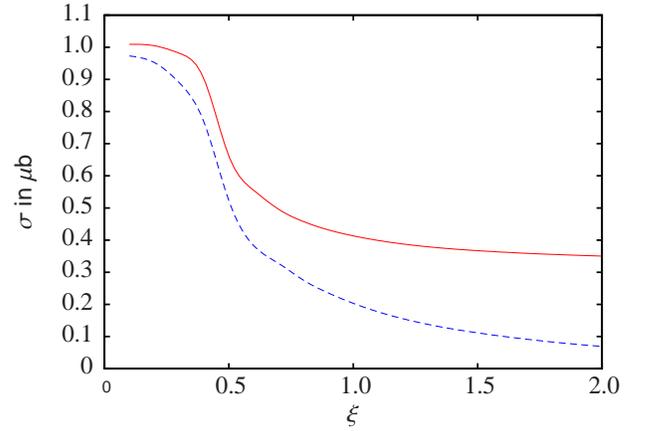}\caption{(Color online) Total cross section as a function of the laser intensity parameter for $\omega=10$ keV and $p^0=250$ MeV$/2\sqrt{1+2\xi_l^2}$ (blue dashed line) for the process $e^+e^-\to \mu^+\mu^-$ in a linearly polarized laser field and corresponding field-free phase average (red solid line). }\label{plot_mu_lp_xidep}
\end{figure}

The dependence of the total cross section on the produced muon's energy is shown in Fig. \ref{plot_mu_lp_spec}. It looks very similar to the differential cross sections shown in Fig. \ref{plot_cp_xi0_5_spec} and Fig. \ref{plot_cp_xi5_spec} for Higgs boson creation. The same similarity in the characteristic features is found for the dependences of the total cross section on the free electron energy $p^0$ (Fig. \ref{plot_mu_lp_pdep}) and on the laser intensity parameter $\xi_l$ (Fig. \ref{plot_mu_lp_xidep}). We again performed a phase average of the field-free cross section, similar to Eq. \eqref{ff_convolution}. As can be seen in Fig. \ref{plot_mu_lp_pdep}, again the dependence of the total cross section on the free electron energy $p^0$ is qualitatively similar in both cases.  The dependence on the laser intensity parameter $\xi$, like in the Higgs boson production case, is in good agreement with the field-free phase average for small $\xi$ and, with our choice of $p^0$, remains dependent on $\xi$ when the latter saturates.

\subsubsection{Required laser parameters}

\begin{table}
 \centering\small
 \begin{tabular}{|r|c|}
 \hline
  Intensity parameter $\xi_l$ &  1\\ \hline 
  Photon energy $\omega$ (eV) &  1\\ \hline 
Laser intensity $I$ (W/cm$^2$) &  $1.8\times10^{18}$ \\ \hline 
Free electron energy $p^0$ (MeV) & 72\\ \hline 
Lorentz factor $\gamma$ &  141 \\ \hline 
Beam radius $\Delta x$ ($\mu$m) &  169\\ \hline 
Pulse duration (ps) &  500\\ \hline 
Pulse power (TW) & 11\\ \hline 
Pulse energy (kJ) &  5.7\\ \hline 
 \end{tabular}
\caption{Minimally required laser parameters for the process ${e^+e^-\to\mu^+\mu^-}$ for the parameter sets considered above for linear polarization.}\label{tab_mu_laserparams}
\end{table}

Like in the case of Higgs boson production, the laser pulse must meet certain requirements in order for the initial electrons to be efficiently accelerated (c.f. Sec. \ref{HiggsParamsSec}). The minimally required pulse parameters in order to cover the whole lepton trajectory for the parameters considered above are listed in Table \ref{tab_mu_laserparams} for linear polarization. {For these parameters, the rate for Compton scattering for the counterpropagating positrons is $R_C\approx5\times10^{10}$ s$^{-1}$, resulting in an energy loss of roughly 2.4 MeV per positron. This is rather small compared to the total collision energy on the order of 250 MeV.} Since the collision energy needed for muon pair production is much smaller than for the production of the much heavier Higgs boson, the required initial electron energy and thus laser pulse power and energy are much smaller than for the former case. Present-day or near-future facilities are in principle capable of reaching them \cite{NIF, Yanovsky, ELI} and thus the process $e^+e^-\to\mu^+\mu^-$ might serve as a proof-of-principle experiment.
 
\subsection{Hadron production}

\begin{table*}
\centering
\small
\begin{tabular}{rccccccc}
\hline
Polarization & circ. & circ. & circ. & lin. & lin. & lin. & lin. \\ \hline
Colliding particles & $e^+e^-$ & $\mu^+\mu^-$ & $\mu^+\mu^-$ & $e^+e^-$ & $e^+e^-$ & $\mu^+\mu^-$ & $\mu^+\mu^-$ \\ \hline
Resonance & $\pi^0$ &  $\Phi$ & $J/\Psi$ & $\pi^0$ & $\Phi$ & $\Phi$ &  $J/\Psi$ \\ \hline
Mass (MeV) & 135 & 1032 & 3097 & 135 & 1032 & 1032 & 3097\\ \hline
Free lepton energy $p^0_\pm$ (MeV) & 50 & 365 & 1095 & 50 & 75 & 298 & 894\\ \hline
Intensity parameter $\xi$ & 0.9 & 1 & 1 & 0.6 & 4.8 & 1 & 1 \\ \hline
Laser intensity $I$ (W/cm$^2$) & $3.5\times10^{18}$ & $1.8\times10^{23}$ & $1.8\times10^{23}$ & $1.6\times10^{18}$ & $9.9\times10^{19}$ & $1.8\times10^{23}$ & $1.8\times10^{23}$ \\ \hline
Lorentz factor $\gamma$ & 98  & 3.5 & 10.4 & 98 & 147 & 2.8 & 8.5\\ \hline
Beam waist $w_0$ ($\mu$m) & 70 & 2.8 & 8.3 & 52 & 564 & 2.3 & 6.8 \\ \hline
Pulse duration (ps) & 130 & 0.2 & 1.8 & 70 & $8.3\times10^3$ & 0.13 & 1.2 \\ \hline
Pulse power (PW) & $0.5$ & 43.2 & 389 & $2\times10^{-3}$ & 1.4 & 10.2 & 30.6 \\ \hline
Pulse energy (kJ) & $70.1$  & 8.7 & 701 & $0.14$ & $1.2\times10^4$ & 1.4 & 37 \\ \hline
\end{tabular}
\caption{Minimally required laser parameters for the production of the listed hadrons in lepton-antilepton collisions. For electron-positron collisions, a free energy of $50$--$75$ MeV has been assumed, from which the required laser intensity parameter has been derived according to the hadron resonance. For muon-antimuon collisions, a laser intensity parameter of $\xi=1$ has been assumed. The laser photon energy has been set to $\omega=1.55$ eV, corresponding to $\lambda=800$ nm.
}\label{hadrons_ELI_e}
\end{table*}

We have seen that the muon-antimuon pair production process and the Higgs boson creation inside a laser field show similar characteristics, such as the reproduction of the field-free cross section in a circularly polarized laser field and, in the case of linear polarization, the behavior of the partial and differential cross sections as well as the dependences of the total cross sections on free lepton energy and laser intensity parameter. This indicates that we can assume similar characteristics in other laser-boosted particle creation processes as well. For example, we may expect that for $s$-channel Higgs boson production at a muon-antimuon collider \cite{Djouadi}, which would be another example of an electroweak process, the underlying principles are also the same and it would be implementable, e.g., with $36$ GeV initial muon-antimuon beams with an 80 ns laser pulse with GJ energy for an intensity parameter $\xi_l=1$ in the linearly polarized case.\\

In this section, we will address the possibility of hadron production at the future Extreme Light Infrastructure without performing the actual calculations. At the ELI, optical laser pulses with wavelengths in the order of $800$ nm and intensities of over $10^{25}$ W/cm$^2$ are envisaged, with peak pulse powers in the 10-PW regime and pulse durations of 200--300 fs, resulting in pulse energies of 200--300 J \cite{ELIbeams}. The intensity $I\sim10^{25}$ W/cm$^2$ would at this wavelength correspond to a laser intensity parameter of $\xi\sim1500$ for electrons and $\Xi\sim7$ for muons. We choose, however, to be more conservative and consider for electrons a preacceleration to $p^0=50$--$75$ MeV which can be obtained, e.g., via laser acceleration and calculate the required laser intensity accordingly. For muons we assume a laser intensity parameter of $\Xi=1$, corresponding to a laser intensity of order $10^{23}$ W/cm$^2$. We now discuss a few exemplary hadrons to be produced in a collider that combines the ELI laser beams with conventional electron-positron or muon-antimuon accelerators, as listed in Table \ref{hadrons_ELI_e}.\\

We consider the production of three different kinds of hadronic resonances, namely neutral pions, the $\Phi$ meson and, lastly, the resonance $J/\Psi$, in the collision of preaccelerated lepton-antilepton beams \cite{footnote1}. As can be seen in Table \ref{hadrons_ELI_e}, circular polarization of the laser field would, due to the large focal area, require large beam energies as compared to linear polarization. However, neutral pions might be producable in electron-positron collisions taking place inside an ELI beam for both polarizations, and the $\Phi$ resonance may well be investigated in a linearly polarized ELI beam with 75-MeV electrons and positrons. The comparison with muon-antimuon collisions shows that, for these rather high-energy processes, the heavier collision particles are favorable for this setup because due to their larger mass, the acceleration to high collision energies requires laser pulses of smaller pulse energy. Thus, heavier hadron resonances can be examined than in electron-positron collisions. Given the broad range of collision energies in the laser-boosted setup, the widths of these resonances will set limits on the production efficiency.

\section{Summary and Conclusion}\label{summarySection}

We have seen in Sec. \ref{HiggsSection} the detailed analytical calculation for the process $\ell^+\ell^-\to HZ$, i.e. the associated production of Higgs and $Z$ bosons in laser-boosted lepton collisions, for both circular and linear polarization. {The collision energy can, by applying an intense laser field, be increased since the particle beam that propagates along with the laser wave will be substantially accelerated by the field. Because the counterpropagating particle beam is virtually unaffected by this effect, this results in an asymmetric collision geometry \cite{footnote2}.} The biggest advantage of linear as compared to circular polarization for our setup lies in the fact that, in order to keep the colliding particles inside the laser wave, smaller spatial extensions of the laser beam and thus a smaller pulse power are necessary. In a circularly polarized laser field, on the other hand, the total cross section and thus the number of produced Higgs bosons is the same as without the laser field, which, depending on the laser intensity parameter $\xi$, can be much larger than the outcome in a linearly polarized laser field. We have also found that, due to the larger mass and thus smaller Lorentz factor for muon-antimuon collisions as compared to electron-positron collisions with the same initial energy, the required laser pulse parameters are in favor of muon-antimuon collisions in such a highly energetic process. {Furthermore, the reduction of the collider luminosity due to the geometry of the laser pulse estimates to be larger for electron-positron collisions, since the required laser pulse needs to have a larger spatial extension than for muon-antimuon collisions. The luminosity loss can be reduced by suitably adjusting the photon energy and beam extensions, as we have seen in Sec. \ref{HiggsParamsSec}.}\\

In Sec. \ref{otherprocesses}, we have shown the calculation for the process $e^+e^-\to\mu^+\mu^-$, i.e. muon pair production in laser-boosted electron-positron collisions, again for circular and linear laser field polarization. This process and the Higgs boson production process discussed earlier, despite being of different natures, show very similar characteristics. \\

In the second part of Sec. \ref{otherprocesses}, we briefly discussed the resonant production of hadrons in laser-boosted lepton collisions. We presume that the general characteristics will be the same as for the two processes investigated in detail. The resonance $\pi^0$ is, in principle, in reach of the ELI laser pulse when combined with electron-positron beams, even in the case of circular polarization. For the creation of particles with higher rest masses, muon-antimuon collisions are favorable. \\

Between the production of muon pairs and resonant $\pi$ meson production in laser-boosted electron-positron collisions and $\Phi$ meson production in muon-antimuon collisions, we identified several possible proof-of-principle experiments that might be realizable in the near to intermediate future. While the experimental demands for the collider scheme considered in this paper are certainly ambitious, it might be interesting to investigate its feasibility in such an experiment and to investigate the opportunities it might be able to offer in future particle colliders.

\begin{acknowledgements}
 Fruitful discussions and useful input by Karl-Tasso Kn\"opfle, Karen Z. Hatsagortsyan, Felix Mackenroth, Kensuke Homma, and Markward Britsch are gratefully acknowledged.
\end{acknowledgements}


\begin{thebibliography}{1}
\bibitem{footnotebeam} {Note that elliptic Gaussian beams, being solutions to the Maxwell equations, can be realized by means of cylindrical focussing lenses. However, the technical details of an actual experimental implementation may require more involved focussing schemes, like, e.g., waveguides consisting of hollow capillaries \cite{capill} or plasma waveguides \cite{waveg} in order to counteract the beam divergence.}
\bibitem{footnote1} {{Note that neutral pions can be produced in electron-positron collisions only under the absorption or production of a photon. The parameters listed in Table III are to be understood as minimal requirements for a certain free electron energy $p^0$ under neglection of said photon's energy.}}
\bibitem{footnote2} {This could be made use of for rendering intrinsically asymmetric collision setups such as electron-proton scattering, where the electron energy is typically much smaller than the proton energy, more symmetric.}
\end{thebibliography}
\end{document}